%% file: rotors.tex
\documentclass[a4wide]{article}
\usepackage[margin=1.5in]{geometry}
\usepackage{times}  
\usepackage{helvet}  
\usepackage{courier}  
\usepackage{url}  
\usepackage{graphicx,dsfont}  
\usepackage{rotating}
\usepackage{amssymb}
\usepackage{amsmath}
\usepackage{graphicx}
\usepackage{url} 

\usepackage{color}
\usepackage{xspace}
\usepackage{blindtext}
\usepackage{subcaption}
\usepackage{tabularx}
\usepackage{tikz}
\usepackage{url}
\usepackage{xcolor}
\usepackage{multirow, rotating}
\usepackage{algorithm,algpseudocode,booktabs,amsmath}
\makeatletter
\renewcommand{\ALG@beginalgorithmic}{\small}
\makeatother

\usepackage{hyperref}
\usepackage{balance}
\usepackage{tikz}
\usetikzlibrary{shapes,decorations,patterns,positioning}
\usetikzlibrary{plotmarks}
\graphicspath{{/}{}}
\newcommand{\ignore}[1]{}

\usepackage{url}
\begin{document} 

\title{Quasi-random Agents for Image Transition and Animation}

\author{
Aneta Neumann\\
Optimisation and Logistics\\
School of Computer Science\\
The University of Adelaide\\
Adelaide, Australia
\and 
 Frank Neumann\\
 Optimisation and Logistics\\
 School of Computer Science\\
 The University of Adelaide\\
 Adelaide, Australia
 \and
 Tobias Friedrich\\
 Chair for Algorithm Engineering\\
 Hasso-Plattner-Institut\\
 Potsdam, Germany
}

\setcounter{secnumdepth}{0}

\maketitle
\begin{abstract}
\begin{quote}
Quasi-random walks show similar features as standard random walks, but with much less randomness. We utilize this established model from discrete mathematics and show how agents carrying out quasi-random walks can be used for image transition and animation. The key idea is to generalize the notion of quasi-random walks and let a set of autonomous agents perform quasi-random walks painting an image. Each agent has one particular target image that they paint when following a sequence of directions for their quasi-random walk. The sequence can easily be chosen by an artist and allows them to produce a wide range of different transition patterns and animations.  
\end{quote}
\end{abstract}

\section{Introduction}

Artificial intelligence techniques have been applied in various scenarios to create or inspire artistic work~\cite{wilson2002information}. We aim at creating artistic art work in the context of image transition and animation. The key idea in this area is to use \emph{randomness} to create interesting processes of visual effects. Previously, random walks have been used in this context~\cite{DBLP:conf/evoW/NeumannAN17}. Random walks and especially the ones biased on the given images can create very interesting effects when it comes to image transition. However, the drawback of this is that it is hard to control by the user/designer of the application. 

This paper presents a new approach for image transition and animation with a controllable amount of randomness. For this, we replace the classical random walk with a so-called \emph{quasi-random walk}. In this model, the decisions of the walker are not chosen uniformly at random, but deterministically. Instead of moving to a random neighbor, the walker visits all neighbors in a fixed order. This model is well-studied in discrete mathematics for investigating how much randomness is required to achieve similar properties of regular random walks, but with less randomness. The quasi-random walk was rediscovered independently several times in the literature. In order of appearance, it has been called 
``Eulerian walker''~\cite{Priezzhev1996},
``edge ant walk''~\cite{WagnerLB99},
``whirling tour''~\cite{DumitriuTW03},
``Propp machine''~\cite{Kleber,CooperSpencer},
``deterministic random walk''~\cite{1DPropp,2DPropp}, and
``rotor-router model''~\cite{FriedrichS10,FriedrichKK15}.
To show the relationship to standard random walks, we mostly use the term ``quasi-random walk'' in the rest of the paper. We will synonymously use ``rotor-router model'' to sometimes emphasize the inner workings.

We use quasi-random walks to design algorithms for image transition and animation~\footnote{Videos are available at \url{https://vimeo.com/user70826513}\label{footnote1}}. In our algorithms, a set of agents perform quasi-random walks where the characteristic of the random walk can easily be determined by a user. 
The process starts with a given starting image (which may be white or blank). Each agent has an image that it paints. It does so by carrying out its quasi-random walk as determined by the router sequence chosen by the user. The $r$ agents perform their quasi-random walks in parallel painting different images (in the case of animation) or the same target image in the case of transition.

The user can easily determine the behavior of each agent by setting the sequence of directions that is followed at every pixel. An example sequence could be (left, down, right, up, left). Each pixel has its current active direction in the sequence.  If the agent visits a pixel $p$, it paints it with the corresponding pixel of its target image. Afterwards, the agent moves to the next pixel as directed by the current active direction at $p$ in its sequence and updates the active direction at $p$ to the next one in the sequence.   

The use of different sequences for the agents allows to create various types of interesting transition and animation processes. Analyzing the images created during the animation process with respect to different artistic features, we show that quasi-random animation using different rotor sequences allows to create animation processes with a wide range of different artistic behaviors.

The remainder of this paper is organized as follows. In the next section, we give a brief introduction into quasi-random walks and present our approach for quasi-random image transition and animation. Afterwards, we examine the wide range of image transitions and animations that can be created using various types of agents. We carry out a feature-based analysis of our animations using artistic features and finish with some  
concluding remarks.

\section{Quasi-random transition and animation}
\label{sec2}
We now describe the concept of quasi-random walks in greater detail and present our method for carrying out quasi-random image transition and animation.

\subsection{Quasi-random walks}
A quasi-random walk is best described on the two-dimensional infinite grid $\mathds{Z}^2$. Imagine a single walker on some arbitrary vertex $(x,y)\in\mathds{Z}^2$. The walker is allowed to move to one of the four neighbors right, down, left, or up, that is, to $(x+1,y)$, $(x,y+1)$, $(x-1,y)$, or $(x,y-1)$. A random
walk would choose independently and uniformly at random one of these four neighbors and move there. In order to use less randomness, the quasi-random walk assigns a permutation of the four directions right, down, left, up to each vertex of the grid. These permutations are fixed initially and deterministically determine the behavior the quasi-random walkers. To store the current status of the permutation, each vertex has a rotor to store which neighbor to visit next. Each time a walker is leaving a vertex, it moves in the direction of the rotor and updates the rotor according to the fixed permutation of the vertex. When the rotor reaches the last position of permutation, it wraps around and starts again at the beginning of the permutation. 
The advantage of the rotor-router model is that it doesn't require randomness and the process is completely determined by the sequence used and the starting position of the rotor.

The simplest rotor sequences are
$(\text{right}, \text{down}, \text{left}, \text{up})$ or
$(\text{right}, \text{left}, \text{down}, \text{up})$.
In both cases the quasi-random walker visits the four
neighbors as uniformly as the standard random walk in expectation.
Note that the first, clockwise rotor sequence is circular while
the second rotor sequence is non-circular.
It has been proven that such quasi-random walkers
behave very similar to the expected behavior of a standard random walk~\cite{2DPropp}. In fact, there is even a slight difference 
between both aforementioned rotor sequences:  The second, non-circular rotor sequence is proven to behave even a bit closer to the expected random walk
than the first, circular rotor sequence.

Following the idea of a ``stack walk''~\cite{HP09}, we generalize the concept of the quasi-random walk and allow not only permutations and rotor sequences of length four, but arbitrary sequences of the four cardinal directions. If some directions occur more frequently than others, this results in a bias of the walker in a particular direction. For example, the rotor sequence $(\text{right, left, up, down, right, left})$ will much more explore horizontally than vertically. This bias implies a deviation from the expected behavior of the standard random walk, but allows for more interesting artistic effects.

\subsection{Related artistic work}

At the begin of the 20th century the well-known artist Paul Klee was part of the famous Bauhaus movement. Klee was a unique university teacher in Weimar and Dessau and an iconic promoter of a theoretical approach to making art.
In the Pedagogical Sketchbook~\cite{PaulKlee1921} the artist shows his innovative approach to artistic expression. One of his important message about art and design is "an active line on the walk, moving freely, without goal."
\begin{figure}[t]    
\centering
\begin{tabular}{cccc}

\subcaptionbox{}{\includegraphics[width = 0.97in]{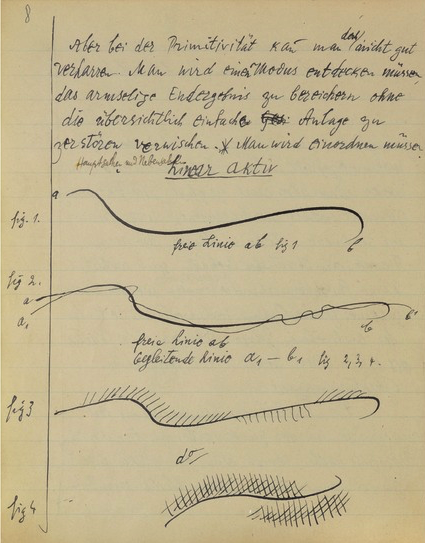}} &
\subcaptionbox{}{\includegraphics[width = 0.955in]{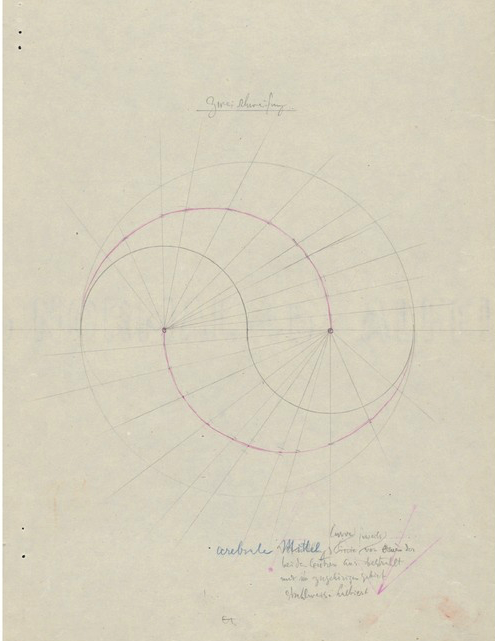}} &
\subcaptionbox{}{\includegraphics[width = 0.81in]{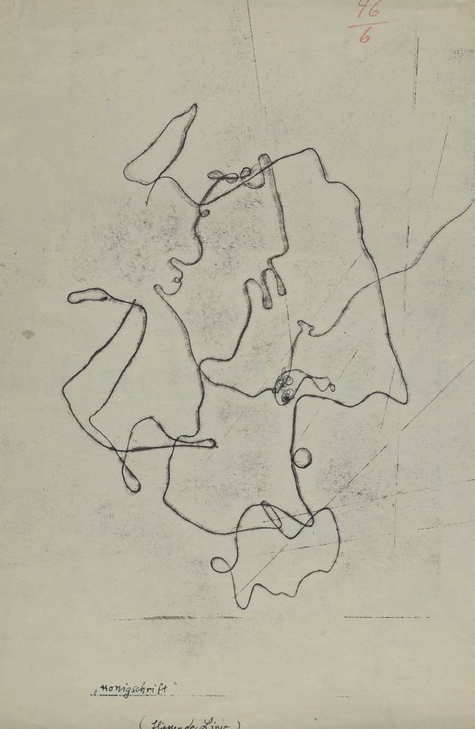}} &

\end{tabular}
\caption{Works of Paul Klee }
\label{PaulKlee}
\end{figure}
He describes the line as the most human mark and characterizes the various types of a line. Figure~\ref{PaulKlee} (a) shows ``the linear line they are subdivided in some modus in contrast to the primitiveness as a free line, a free line with accompanies of lines''. In (b), we see the drawing of ``construction with middle curve from two centers radiated and in associated region shaft-wide halved''. Finally, in (c), we see the drawing with the title Honigschrift as ``a flowing line''. Paul Klee's notes include 3900 pages and demonstrate in multiple ways how a point can become a line and the line the plane can result in ``free irregularity''.
Quasi-random walks essentially create lines in the space where they perform the walk and we take this as an inspiration for designing our transition and animation methods.

\subsection{Our approach}

We take the technical behavior of quasi-random walks and combine it with Klee's idea to create artistic work.
\emph{Quasi-random animation} follows the idea of the rotor-router model in order to produce artistic ideas by agents carrying out quasi-random walks. The pseudo-code for quasi-random animation is given in Algorithm~\ref{alg:qranimation}.

We consider a set of $r$ agents where each of them has the goal to paint one particular image $I^k$. 
Each agent $k$, $1\leq k \leq r$, works with a sequence $S^k$ which consists of entries from $\{\text{right}, \text{down}, \text{left}, \text{up}\}$. For example, we could have $S^k=(\text{right, left, down, up})$ as in the standard rotor-router model, but also sequences such as $S^k = (\text{right, left, up, down, right, left})$ which are not symmetric with respect to the different directions. Using such asymmetric sequences creates a bias in the walk performed by the agent and leads to various interesting effects dependent on the choice of $S^k$. At each time step each agent $k$ moves from its current position $P^k$ to one of its pixel neighbors $\hat{P}^k \in N(P^k)$ where the direction $S^k(c(P^k))$ is determined by the current active pointer at position $P^k$ (wrapping around at the border of the image). 
Afterwards the current image $X$ gets updated by setting the pixel value at position $\hat{P}^k$ to the value of image $I^k$ at position $\hat{P}^k$. The move of agent $k$ is completed by increasing the sequence counter $c^k(P^k)$ in a modulo fashion and updating the current position $P^k$ to $\hat{P}^k$.

\emph{Quasi-random image transition} is a special case of quasi-random animation where all $I^k$ are set to one particular target image $I$.

\begin{algorithm}[t]
\caption{\textsc{Quasi-random Animation}}
\label{alg:qranimation}
\begin{algorithmic}[1]
 \Require{Start image $Y$ of size $m \times n$. For each agent k, $1 \leq k \leq r$, an image $I^k$ of size $m \times n$, sequence $S^k$ and position counters $c^k(i,j) \in \{0, \ldots, |S^k|\}$, $1 \leq i \leq m$, $1 \leq j \leq n$.}
\State $X \gets Y$
\For{each agent $k$, $1 \leq k \leq r$}
\State choose $P^k \in m \times n$ and set $X(P^k) := I^k(P^k)$.
\EndFor
\State $t \gets 1$
\While{ $(t \leq t_{\max})$}
\For {each agent $k$, $1 \leq k \leq r$}
\State Choose $\hat{P}^k \in N(P^k)$ according to $S_k(c(P^k))$.
\State $X(\hat{P}^k) \gets I^k(\hat{P}^k)$
\State $c^k(P^k) \gets (c^k(P^k) +1) \mod |S^k|$.
\State $P^k \gets \hat{P}^k$.
\EndFor
\State $t \gets t + 1$
\EndWhile
\end{algorithmic}
\end{algorithm}

The key aspects when using this algorithm to create interesting transitions and animations is the number of agents, the images that they are using for painting and the router sequence for each agent. The choice of the router sequence $S^k$ determines the random walk behavior of the agent $k$. A sequence can be an arbitrary sequence of directions. We mainly study sequences that are rather short in this paper. The reason is that this already gives a large variety of different effects. Furthermore, short sequences are user friendly as they can be easily adapted and understood by the user. The potential of longer sequences (partially explored in our feature-based analysis) allows to design more fine grained animations. In addition, the approach may be extended to have for each pixel its own sequence of directions which would allow to resemble the behavior of biased random walks used for example in image segmentation~\cite{DBLP:journals/pami/Grady06}.

A sequence of length $4$ containing each direction exactly once such as (right, down, left, up) corresponds to the classical roter-router model leading to a cover time that is close to the one of a classical random walk which moves at each to a neighboring pixel selected uniformly at random. However, the setting leads to a much smoother transition and animation being generated as it avoids the unbalance in terms of the different directions that occurs in the classical random walk.
An unbalanced sequence such as (right, down, left, up, right) where each direction does not appear the same number of times, introduces a bias into the direction that appears more often. Let $r$ be the total sequence length and $r_i$ the number of times direction $i$ appears in this sequence. Assume that the agent has done $M$ steps, where $M$ is large. Then the number of steps it has taken into direction $i$ is roughly 
$M \cdot \frac{r_i}{r}.$

This means that the sequence length $r$ and the number of appearances $r_i$ of direction $i$ can be used to create arbitrary fractions of moves into the different directions. Given that each agent is working with its own sequence and therefore its own fractions of the different directions, complex behavior can be created by using agents with different types of sequences.

\section{Results for Quasi-random image transition}
\label{sec3}

\input{exptrans.tex}

\section{Results for Quasi-random Animation}
\label{sec4}

\input{expan.tex}

\section{Feature-based Analysis}
\label{sec5}

We now analyze the different animation processes obtained by our quasi-random animation algorithm. As the process creates different images over time, we measure the images created with respect to different artistic features. We consider features from the literature that have been used to measure artistic images~\cite{COMPAESTH:COMPAESTH05:169-176,matkovic2005global,hasler2003measuring,DBLP:conf/chi/ReineckeYMMZLG13,DBLP:journals/swevo/HeijerE14,DBLP:conf/evoW/NeumannAN17} in order to enable an objective evaluation.

\begin{figure*}[!t]     
\centering
\begin{tabular}{cccccccc}

{\includegraphics[width = 1.3in]{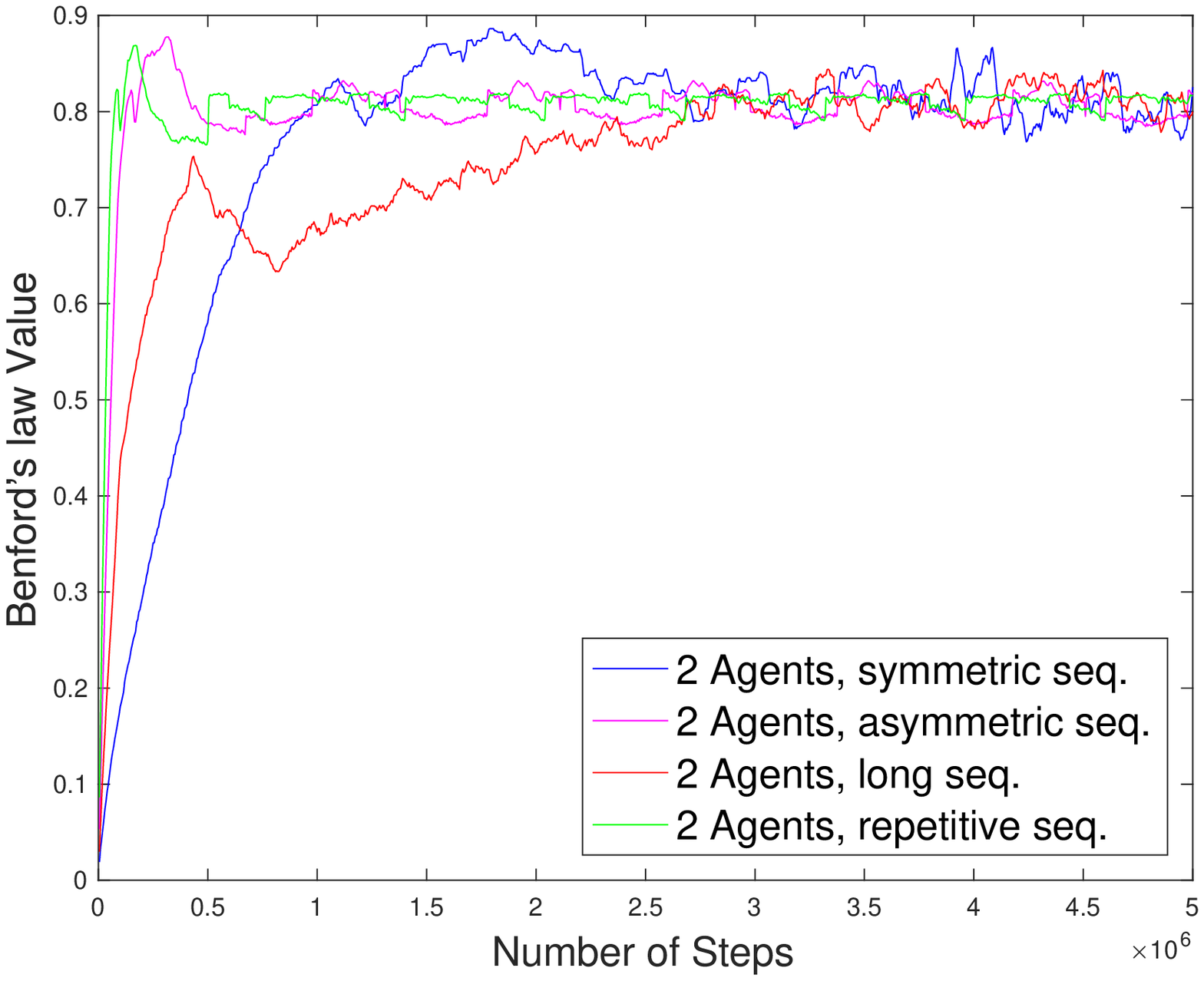}} & 

{\includegraphics[width = 1.3in]{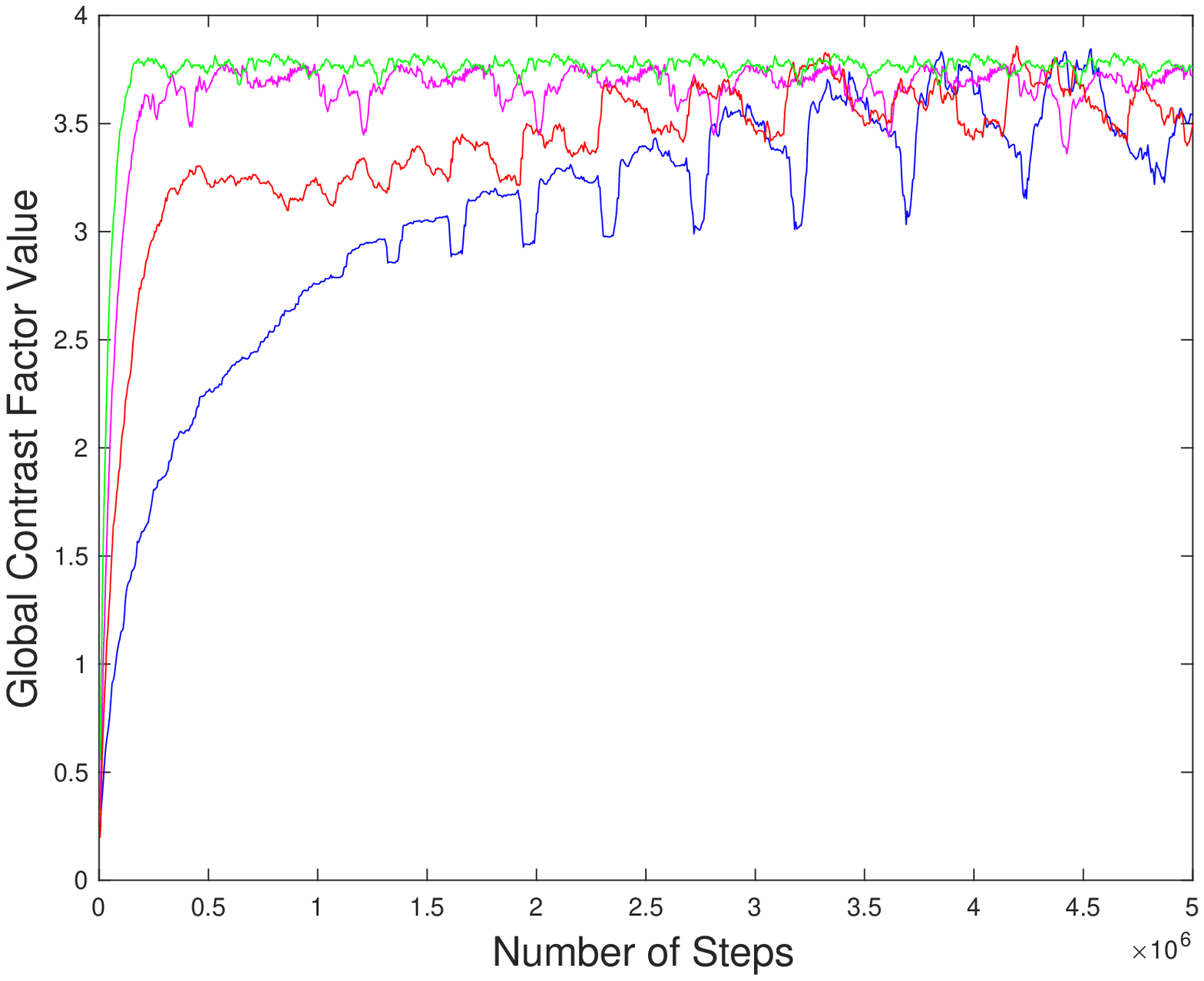}} &

{\includegraphics[width = 1.3in]{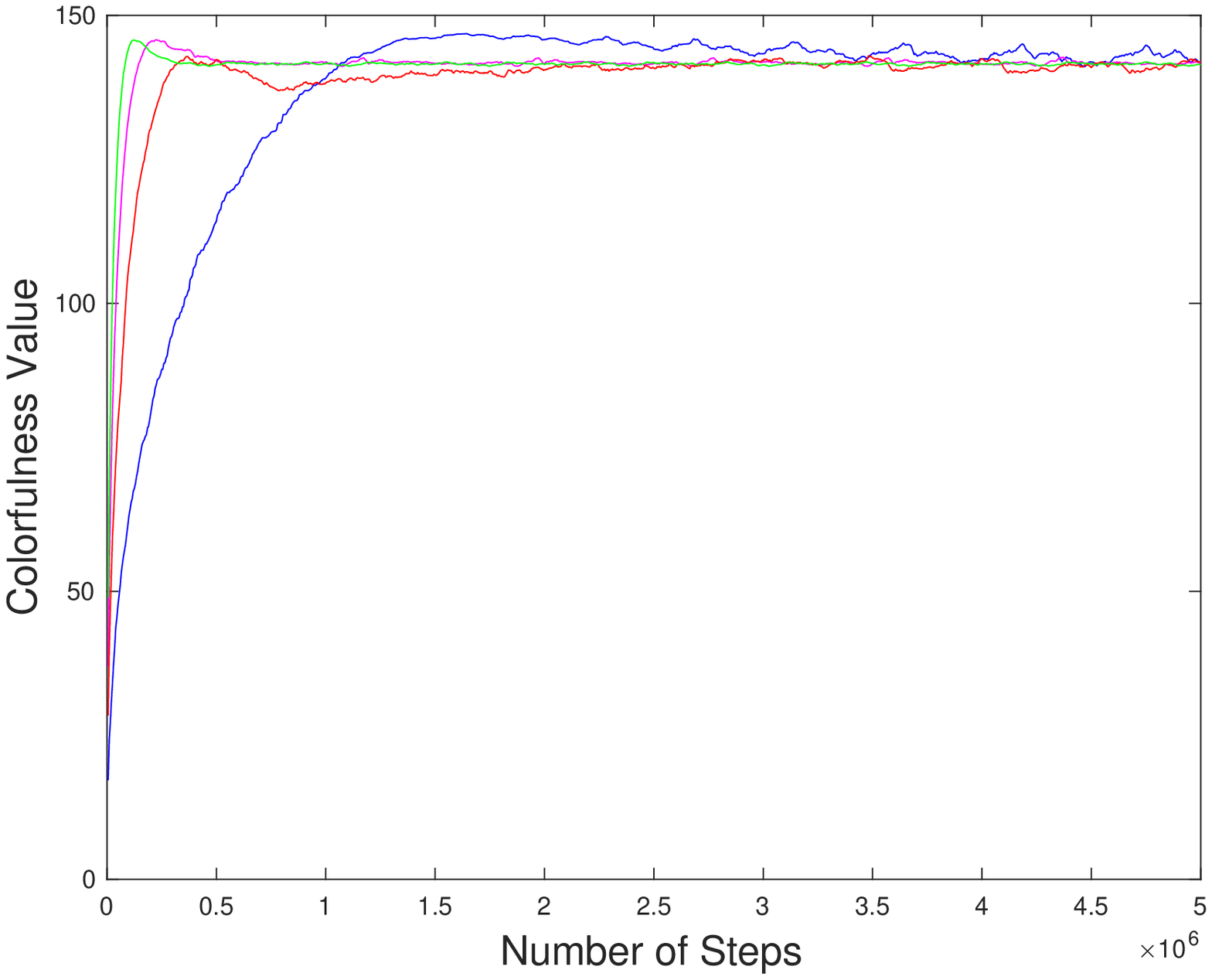}} &

{\includegraphics[width = 1.3in]{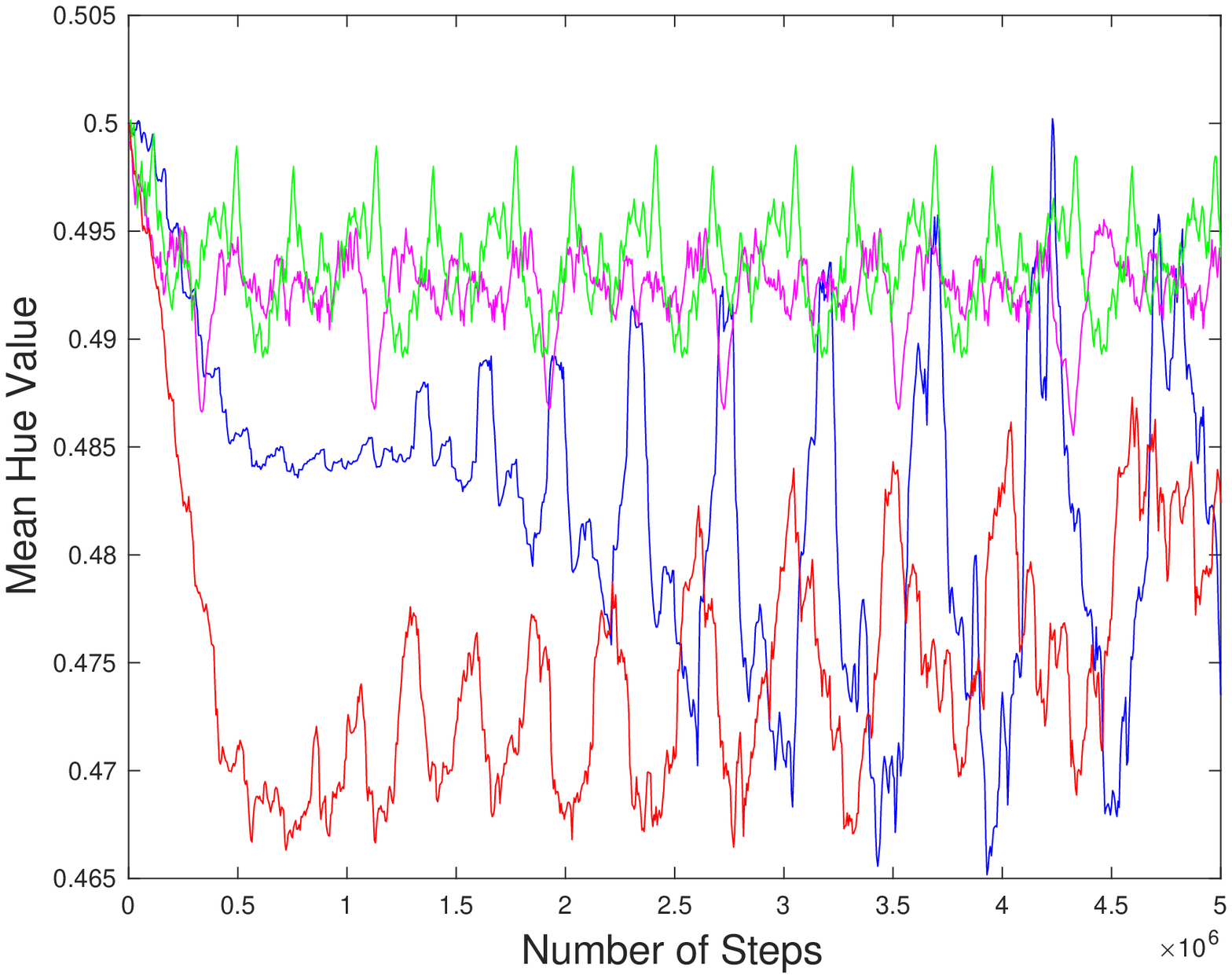}} \\

\subcaptionbox*{(a) Benford's law}{\includegraphics[width = 1.3in]
{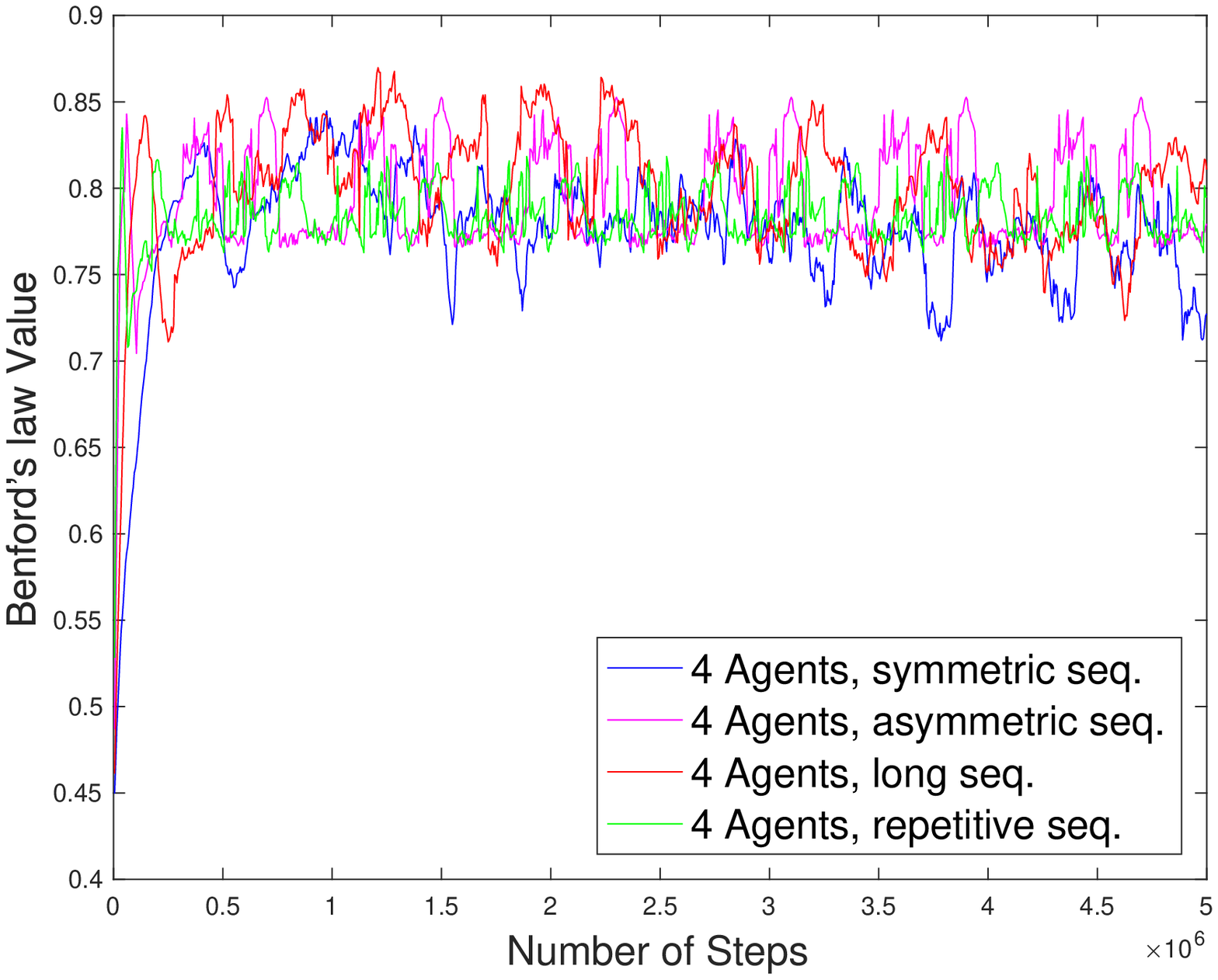}} &
\subcaptionbox*{(b) Global Contrast Factor}{\includegraphics[width = 1.3in]{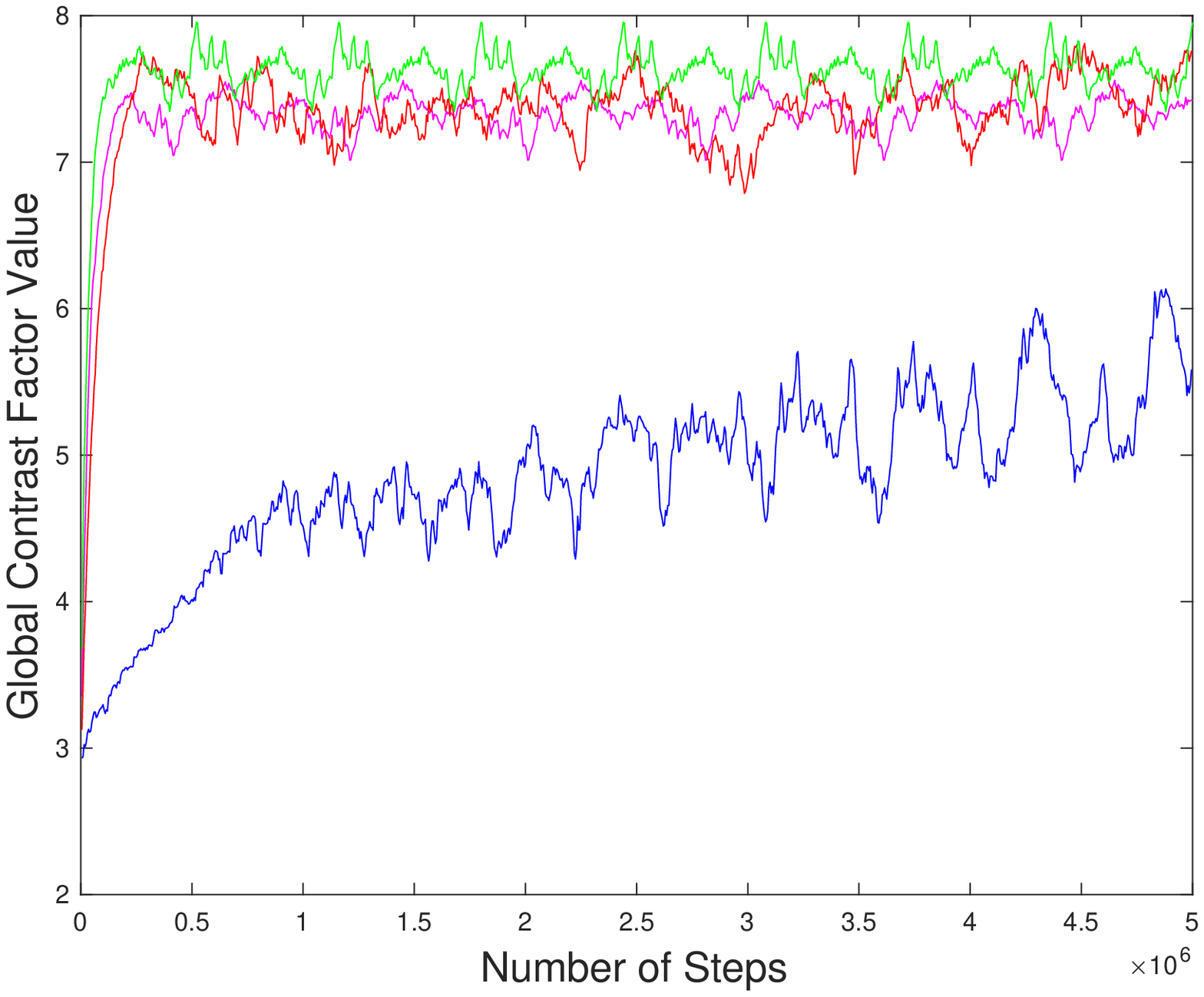}} &
\subcaptionbox*{(c) Colorfulness}{\includegraphics[width = 1.3in]{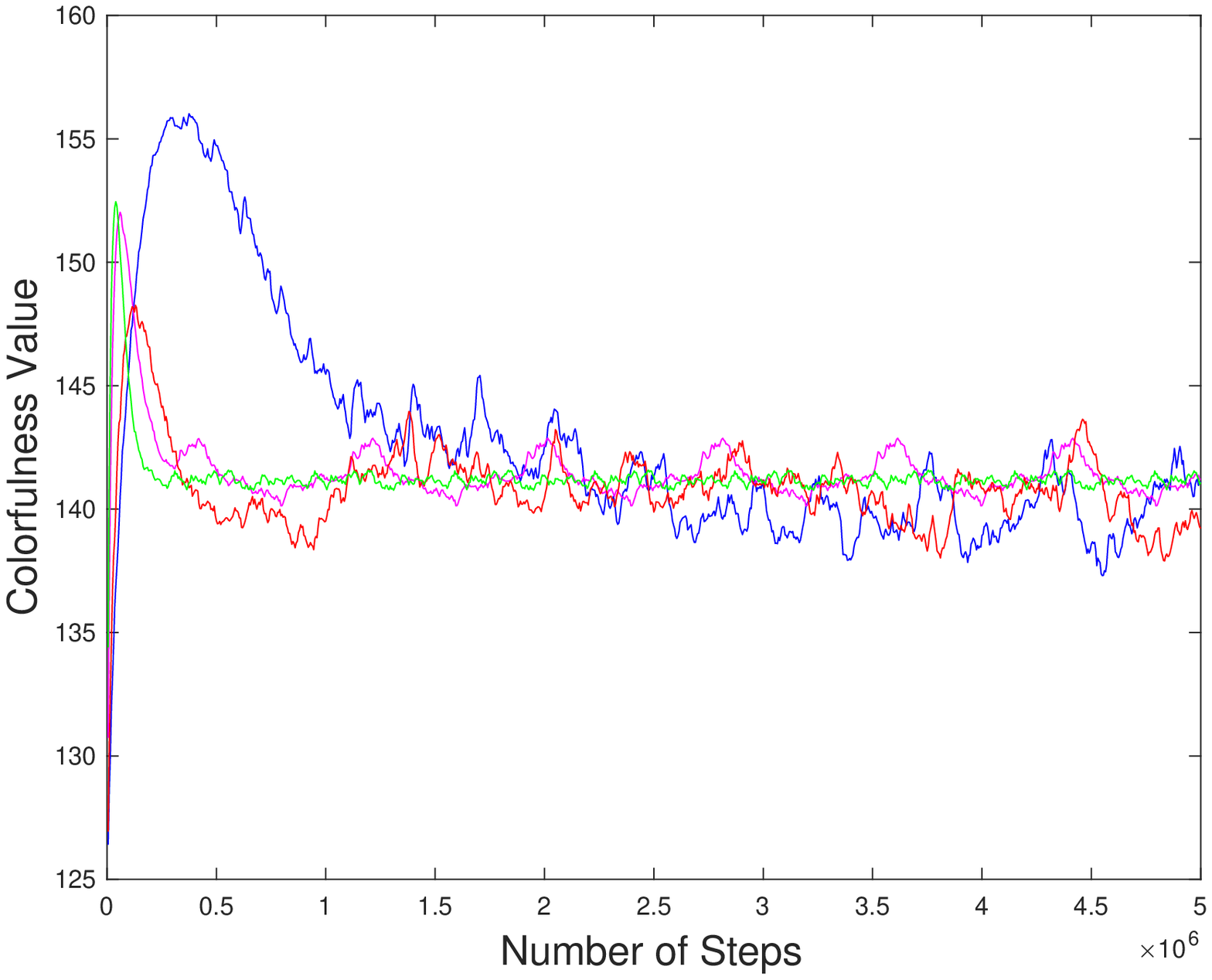}} &
\subcaptionbox*{(d) Mean Hue}{\includegraphics[width = 1.3in]{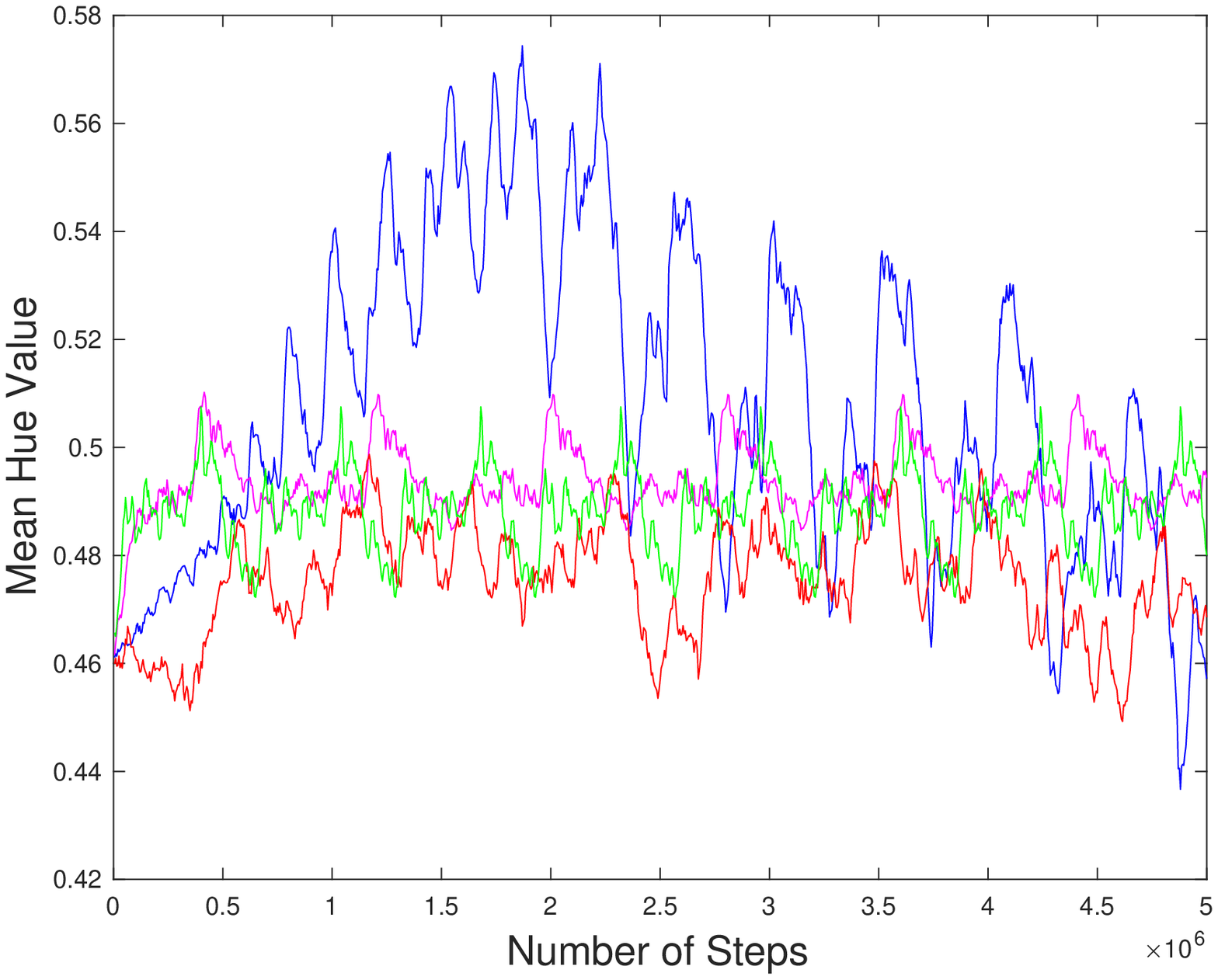}} \\

\end{tabular}
\caption{Features during quasi-random animation}
\label{Plots}
\end{figure*}

For our analysis, we examine all animation settings with $2$ and $4$ agents studied in the previous section. Furthermore we investigate the following additional settings for $2$ agents: In the \emph{long sequences} experiment, we consider $S^1$ = (right, down, left, up), $S^2$ = (right, down, left, up, right, down, left, up, right, down, left, up, right, down, left, up, right, up) and in the \emph{repetitive sequences} experiment we use
$S^1$ = (up, right, right, right), $S^2$ = (down, right, right, right).
For $4$ agents we consider the following two additional settings:
In the \emph{long sequences} experiment, we use
$S^1=S^3$ = (right, down, left, up), $S^2=S^4$ = (right, down, left, up, right, down, left, up, right, down, left, up, right, down, left, up, right, up). For the \emph{repetitive sequences}  experiment, we use $S^1=S^3$ = (up, right, right, right), and $S^2=S^4$ = (down, right, right, right). 
The results for all animations according to the features \emph{Benford's law}, \emph{Global Contrast Factor}, \emph{Colorfulness}, \emph{Mean Hue} are shown in Figure~\ref{Plots}. Here the top row shows the results for animations with $2$ agents whereas the bottom row shows the results for $4$ agents.

Subfigure (a) shows the results of the different animations for the feature \emph{Benford's law}~\cite{benford1938law,hasler2003measuring}. It can be observed that the curve for $2$ agents with symmetric sequences between 1\,500\,000 and 2\,000\,000 steps achieve the highest value. High values mean that results are less natural. Furthermore, the 2 agents with long sequences show the most natural results, as the values in comparison to the other agents are the smallest. During the animation, the values increase slowly. The values for $2$ agents with repetitive and asymmetric sequences show a more stable behavior during the animation. 
The feature values for the animations that involves the use of 4 agents with long and asymmetric sequences show regularly appearing differences in values during the animation. They also have the overall the highest feature values among all animations. This means that those animations are less natural and explains the chaotic appearance of the animation.
Finally, it can be observed that the curve for $4$ agents with asymmetric sequences is clearly lower than those obtained for $4$ agents with symmetric sequences.

In Subfigure (b), the results for the feature \emph{Global Contrast Factor} show a great difference between symmetric sequences and all other choices. \emph{Global Contrast Factor} measures the richness of contrast and corresponds to the human perception of contrast in terms of visual attention. 
\cite{matkovic2005global} show that humans prefer higher scores for \emph{Global Contrast Factor} if they are looking at images.
High values during the animation can be observed for the animations with 2 and 4 agents using repetitive sequences. 
For the animation with $2$ and $4$ agents with symmetric sequences, a characteristic zig-zag curve is obtained where the value slowly increases over time. In contrast to this, the animations with $2$ and $4$ agents using asymmetric, long and repetitive sequences have less variation in term of the feature values.

Subfigure (c) shows the results of the different animations for the feature \emph{Colorfulness}. \cite{hasler2003measuring} computed perceived colorfulness to evaluate the effect that processing of the natural images has on their colour. \cite{DBLP:conf/chi/ReineckeYMMZLG13} have used measurements of perceived colorfulness of website screenshots to developed computational models. It can be observed that
all curves of the \emph{Colorfulness} feature obtain values of around $140$-$145$ after the animation has run for 1\,000\,000 steps. 
Additionally, all curves have a very steady behavior except for the $4$ agents with symmetric sequences where we can observe regular changes over time. Such changes make the process more interesting as it shows a larger variety within the animation. 

In Subfigure (d), the results for the feature \emph{Mean Hue} are shown. The animation with $4$ agents using symmetric sequences reveal the largest variety in terms of this feature value. It can be observe that the curve is alternating between high and low values after approx. 2\,500\,000 steps. 
This expresses the chaotic character of those agents movement during the animation. In contrast to this, the curve obtained during the animation for $2$ agents with long sequences has a zig-zag pattern with tendency to increasing those values during the animation. On the other hand, all agents with repetitive and asymmetric sequences tend to have 
a continuous and regular behavior during the quasi-random animation.

\section{Conclusions}
Quasi-random methods allow to create interesting random behaviors and can easily be determined by a user who only has to set a few parameters within a quasi-random algorithm.
We have presented a new approach to carry out image transition and animations. Our approach builds on theoretical insights of quasi-random walks and generalizes this concepts to multiple agents performing quasi-random walks for image transition and animation. The approach allows to create different forms of transition and animation by determining a sequence for each agent to be used in the quasi-random walk. Choosing these sequences is very easy and allows an artist to experiment with different complex behaviors resulting in interesting ways of carrying out transition and animation. Our feature-based analysis shows that the resulting animations exhibit quite different behavior in terms of important artistic features.

\section{Acknowledgment}
This research has been supported by the Australian Research Council (ARC) through grants DP140103400 and DP160102401.

\bibliographystyle{abbrv}

\bibliography{neumann,references,router}

\end{document}

%% file: exptrans.tex
\begin{figure}[t]     
\centering
\begin{tabular}{cccc}
\subcaptionbox*{(a)}{\includegraphics[width = 1.59in]{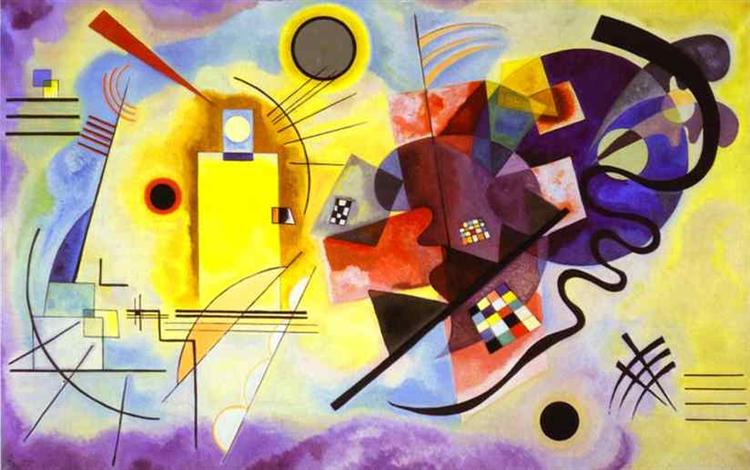}}&
\subcaptionbox*{(b)}{\includegraphics[width = 1.437in]{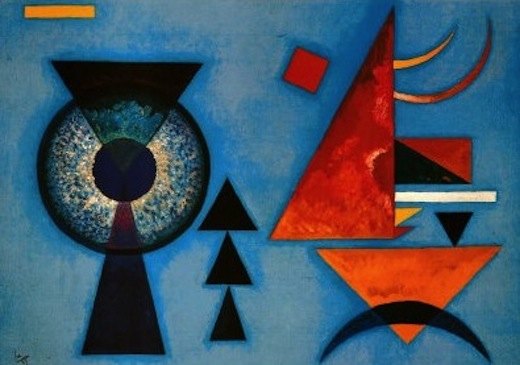}}\\
\subcaptionbox*{(c)}{\includegraphics[width = 0.975in]{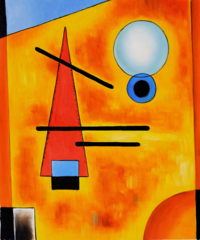}} &
\subcaptionbox*{(d)}{\includegraphics[width = 1.58in]{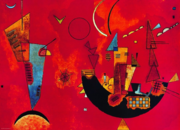}} \\

\end{tabular}
\caption{Starting and target images}
\label{source_and_targets}
\end{figure}

We now show our results for quasi-random image transition where all agents paint the same image $I$.
We use the pairs of images shown in Figure~\ref{source_and_targets} for our experiments.

 To illustrate the effect of the diverse methods presented in this paper, we consider the Yellow-Red-Blue, 1925 (a) and the Cool, 1927 by Wassily Kandinsky (c) as a starting image. We consider the Soft Hard, 1927 (b) and the With and Against, 1929 by Wassily Kandinsky (d) as a target image for image transition.  
In the experiments, each agent performs 5\,000\,000 steps and the images are of size 400\,x\,400 pixels. This setting holds for all experiments reported on in this paper.
We should mention that the use of other starting and target images would show the same algorithmic behavior as the performance of the agents in Algorithm~\ref{alg:qranimation} is independent of the choice of the images. However, using different images allows to create different artistic effects and we will use different sets of images to illustrate a wide range of effects we observed.

\subsubsection{1 Agent}

\begin{figure*}[!t]     
\centering
\begin{tabular}{cccccccc}
\subcaptionbox*{}{\includegraphics[width = 1.225in]{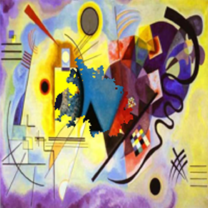}} & 
\subcaptionbox*{}{\includegraphics[width = 1.225in]{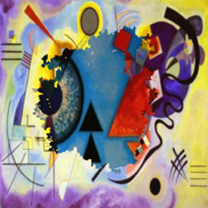}} & 
\subcaptionbox*{}{\includegraphics[width = 1.225in]{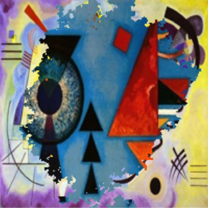}} & 
\subcaptionbox*{}{\includegraphics[width = 1.225in]{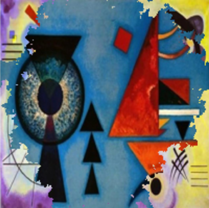}} \\

\vspace{-0.3cm}
\subcaptionbox*{}{\includegraphics[width = 1.225in]{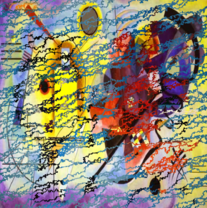}} & 
\subcaptionbox*{}{\includegraphics[width = 1.225in]{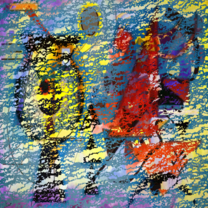}} & 
\subcaptionbox*{}{\includegraphics[width = 1.225in]{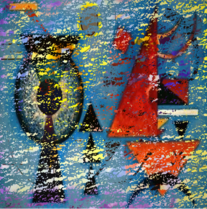}} & 
\subcaptionbox*{}{\includegraphics[width = 1.225in]{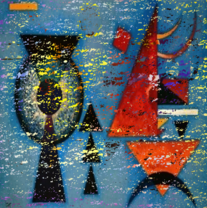}} \\ 
\end{tabular}
\caption{Image transition with $1$ Agent. Top row shows symmetric sequences (images after~\,100\,000, 500\,000, 2\,000\,000 and 3\,500\,000 steps).
Bottom row shows asymmetric sequences (images after~\,100\,000, 200\,000, 300\,000 and 500\,000 steps)}
\label{One_Walker}
\end{figure*}

In our first set of experiments, we consider $1$ agent carrying out the image transition. For the first experiment, we consider the sequence $S^1$=(right, down, left, up) which resemble a standard roter-router model.
The results in Figure~\ref{One_Walker} (top row) shows the quasi-random image transition process in a circle-patch style transforming image $(a)$ into image $(b)$. As we set the starting point of router at the middle of a given image $(a)$, we observe the movement of router circulated in right direction over the image. The quasi-random image transition produces a visual pleasing results with bridging the parts of both images in a new abstract composition. Compared to image transition using classical random walks \cite{DBLP:conf/evoW/NeumannAN17}, it produces much smoother images.

In the second experiment, we expand our approach to non standard roter-router settings and use an asymmetric sequence for the transition. We add only one additionally position to the sequence which introduces a bias into the added direction.
For the experiment, we consider the sequence $S^1$= (right, down, left, up, right).
Figure~\ref{One_Walker} (bottom row) shows the results. In contrast to the previous setting, the behavior of the quasi-random image transition leads to horizontal stripes over the whole image. These stripes gather over time leading to interesting cumulative effects with the target image finally appearing. It also results in a much faster transition process. This is due to the additional direction "right" in the chosen sequence.

\subsubsection{2 Agents}

\begin{figure*}[!t]     
\centering
\begin{tabular}{cccccccc}

\subcaptionbox*{}{\includegraphics[width = 1.225in]{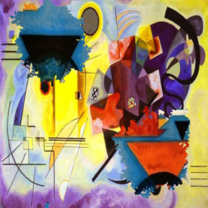}} & 
\subcaptionbox*{}{\includegraphics[width = 1.225in]{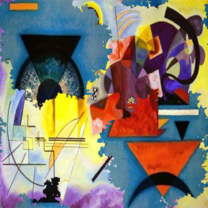}} & 
\subcaptionbox*{}{\includegraphics[width = 1.225in]{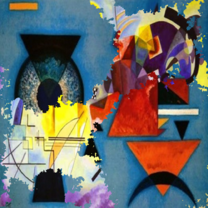}} & 
\subcaptionbox*{}{\includegraphics[width = 1.225in]{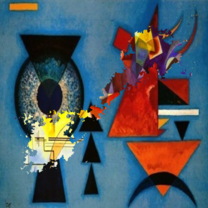}} \\ 
\vspace{-0.3cm}

\subcaptionbox*{}{\includegraphics[width = 1.225in]{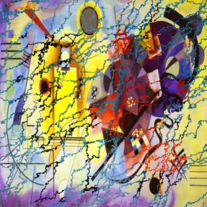}} & 
\subcaptionbox*{}{\includegraphics[width = 1.225in]{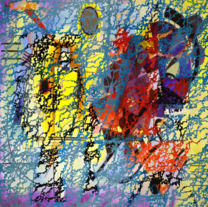}} & 
\subcaptionbox*{}{\includegraphics[width = 1.225in]{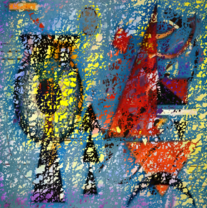}} &
\subcaptionbox*{}{\includegraphics[width = 1.225in]{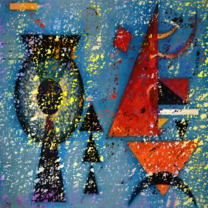}} \\ 

\end{tabular}
\caption{Image transition with $2$ Agents. 
Top row shows symmetric sequences (images after~\,200\,000, 600\,000, 1\,000\,000 and 1\,900\,000 steps).
Bottom row shows asymmetric sequences (images after~\,30\,000, 80\,000, 140\,000 and 210\,000 steps)}
\label{Two_Walkers}
\end{figure*}

We now consider the combination of $2$ agents performing the image transition.  
The starting positions of the $2$ agents are chosen as $[m/4, n/4]$ and $[3m/4, 3n/4]$. We consider the sequences $S^1$ = (right, down, left, up) and $S^2$ = (up, left, down, right) for the agents. The results in Figure~\ref{Two_Walkers} (top row) show two parts of the target image around the agents starting positions emerging. The process of quasi-random image transition with an arrangement of $2$ agents appears more dynamic than the one with $1$ agent using a symmetric sequence.

Finally, we study the behavior of $2$ agents with asymmetric sequences for image transition.
We consider the sequences $S^1$ = (right, down, left, up, right) and $S^2$ = (up, left, down, right, up).
The results in Figure~\ref{Two_Walkers} (bottom row) show that the quasi-random image transition progresses similar to the $1$ agent approach with an asymmetric sequence. In addition, there are vertical stripes "walking/running" from left to right due to the second agent. The whole transition occurs by horizontal and vertical quasi-random walks on the image.

%% file: expan.tex
We now present results obtained for quasi-random animation. We showcase the results using interesting and characteristic images of the considered animation process. The different animation videos are available online~\ref{footnote1}. Furthermore, the feature-based analysis in the next section provides an analysis over the whole animation process with respect to artistic features. 
\subsubsection{2  Agents}

\begin{figure*}[!t]     
\centering
\begin{tabular}{cccccccc}

\subcaptionbox*{}{\includegraphics[width = 1.225in]{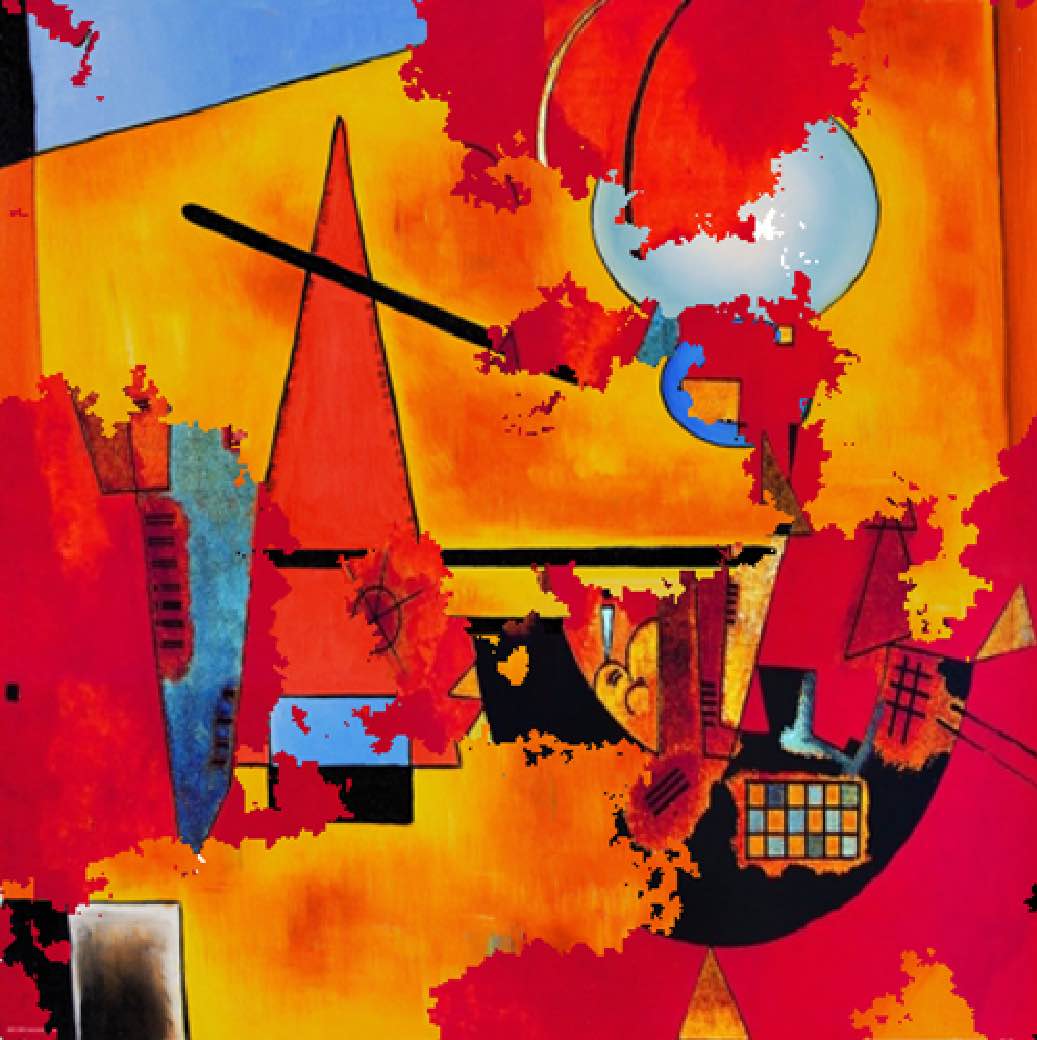}} & 
\subcaptionbox*{}{\includegraphics[width = 1.225in] {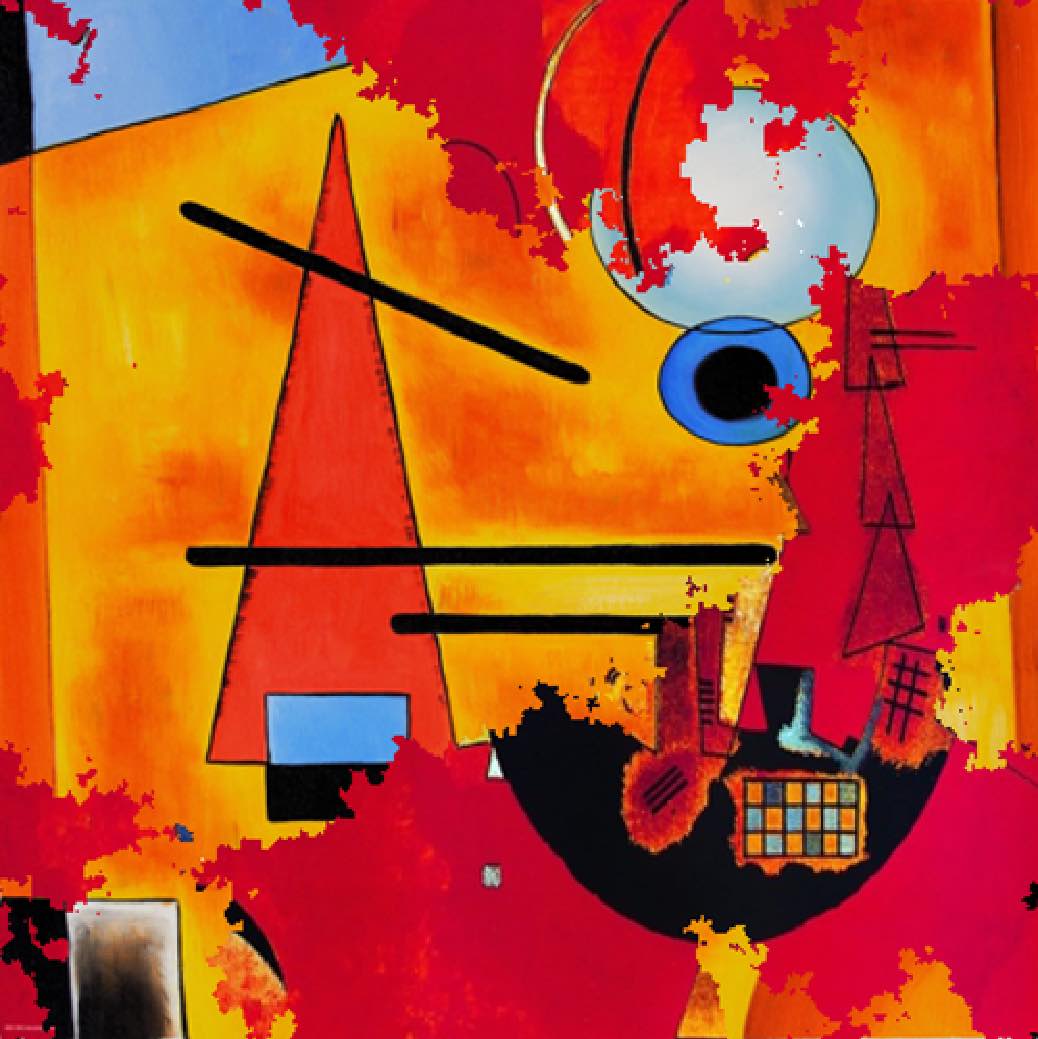}} & 
\subcaptionbox*{}{\includegraphics[width = 1.225in]{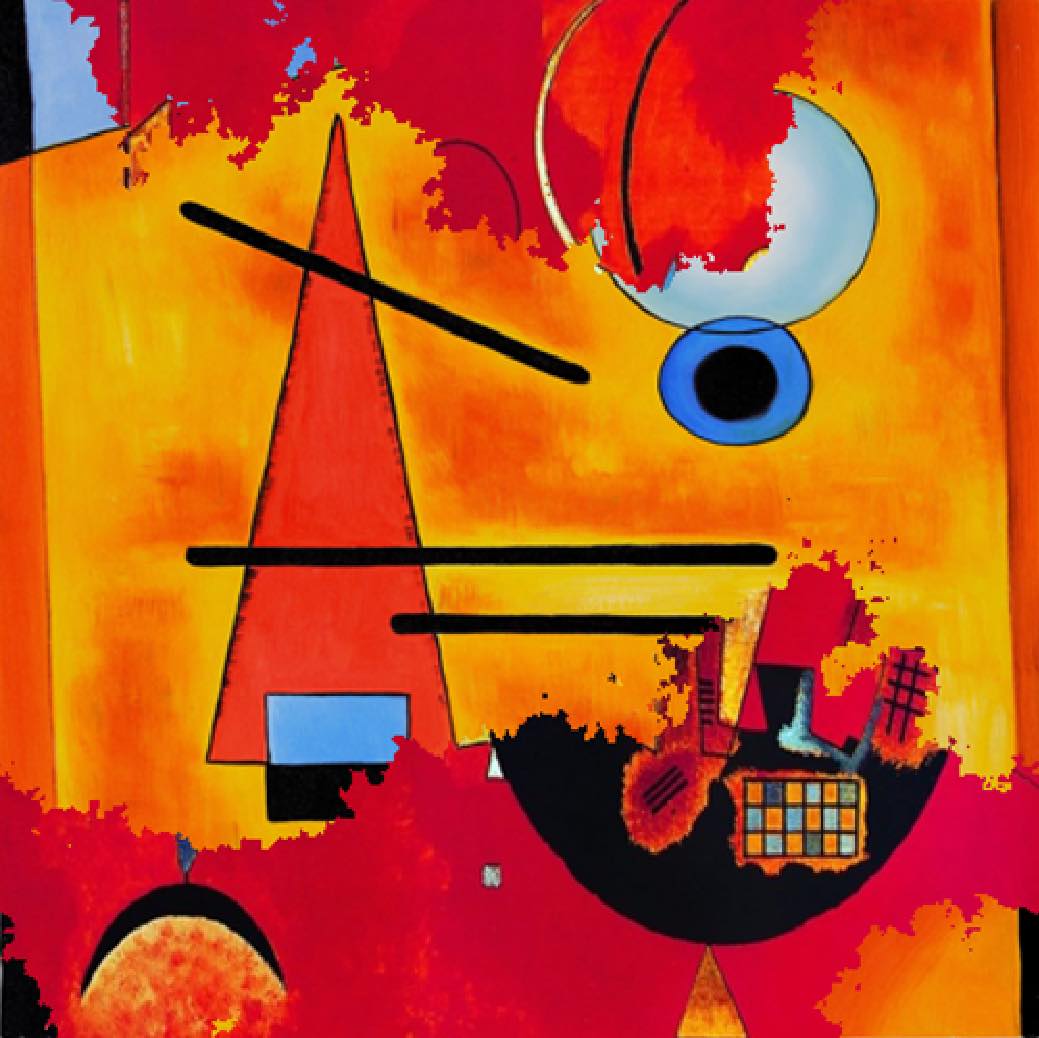}} & 
\subcaptionbox*{}{\includegraphics[width = 1.225in]{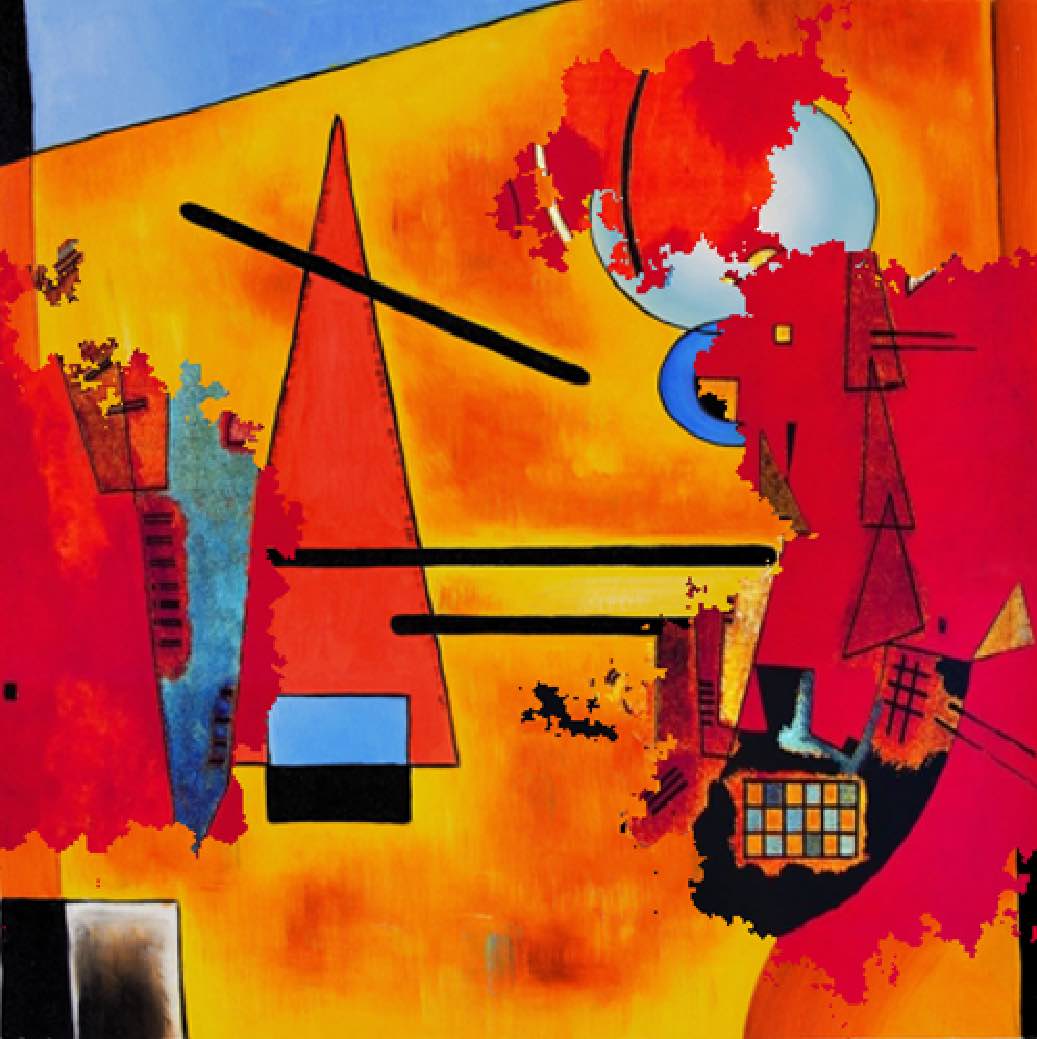}} \\  
\vspace{-0.3cm}

\subcaptionbox*{}{\includegraphics[width = 1.225in]{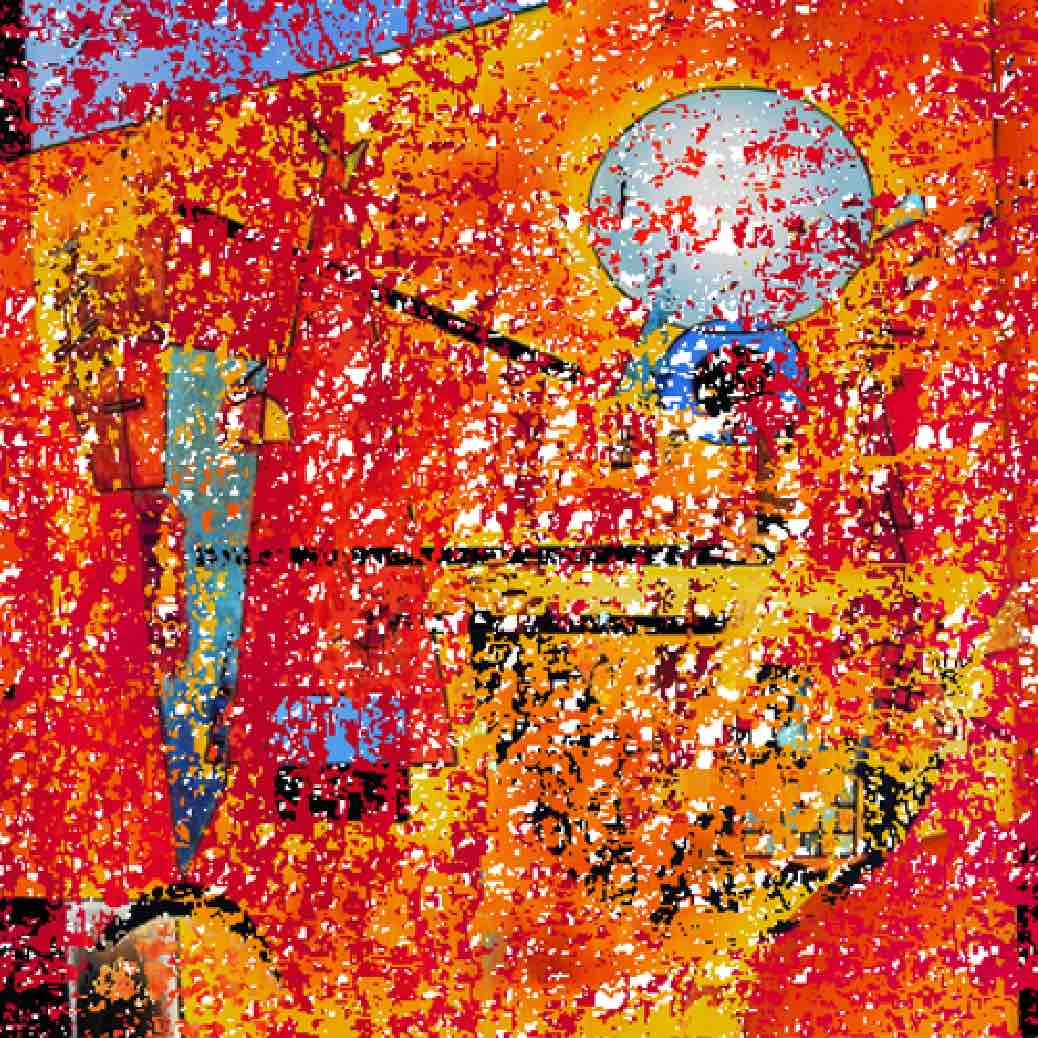}} & 
\subcaptionbox*{}{\includegraphics[width = 1.225in]{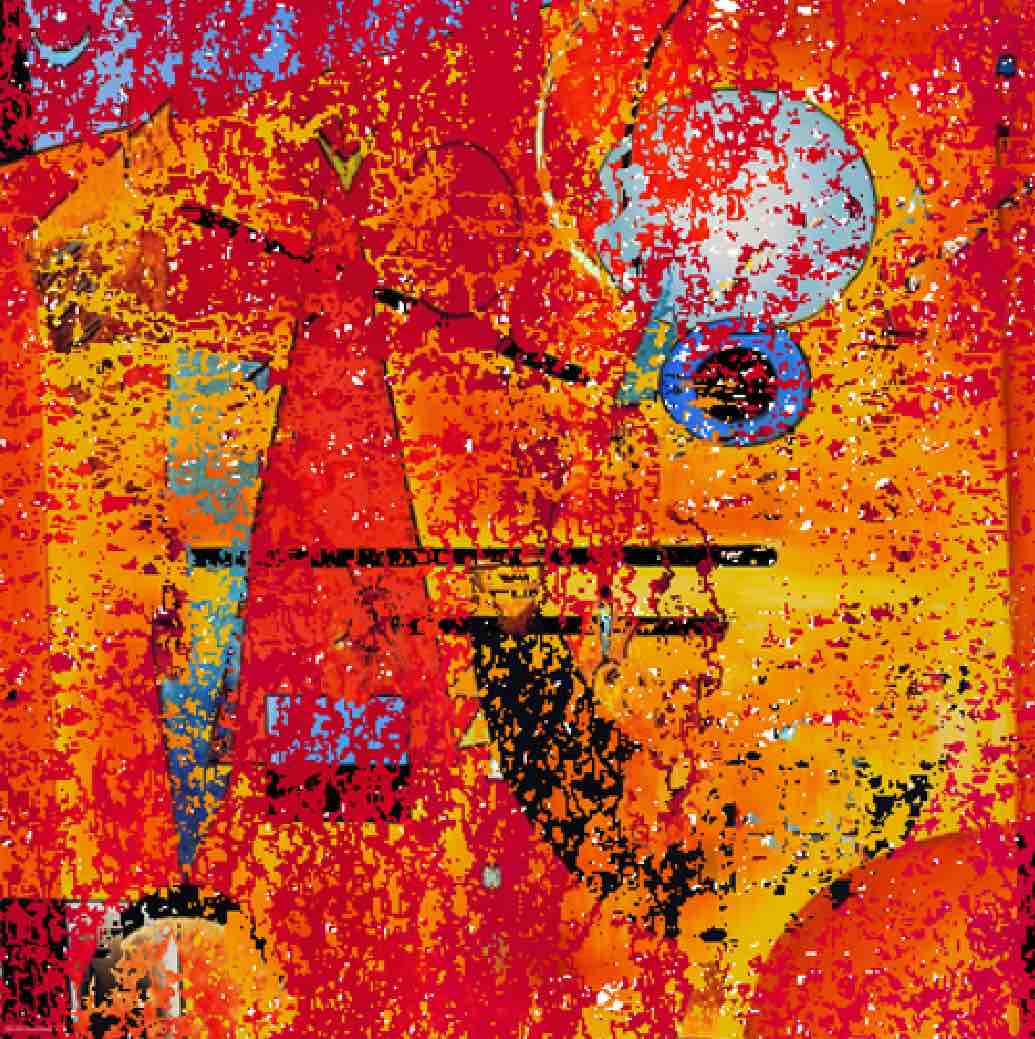}} & 
\subcaptionbox*{}{\includegraphics[width = 1.225in]{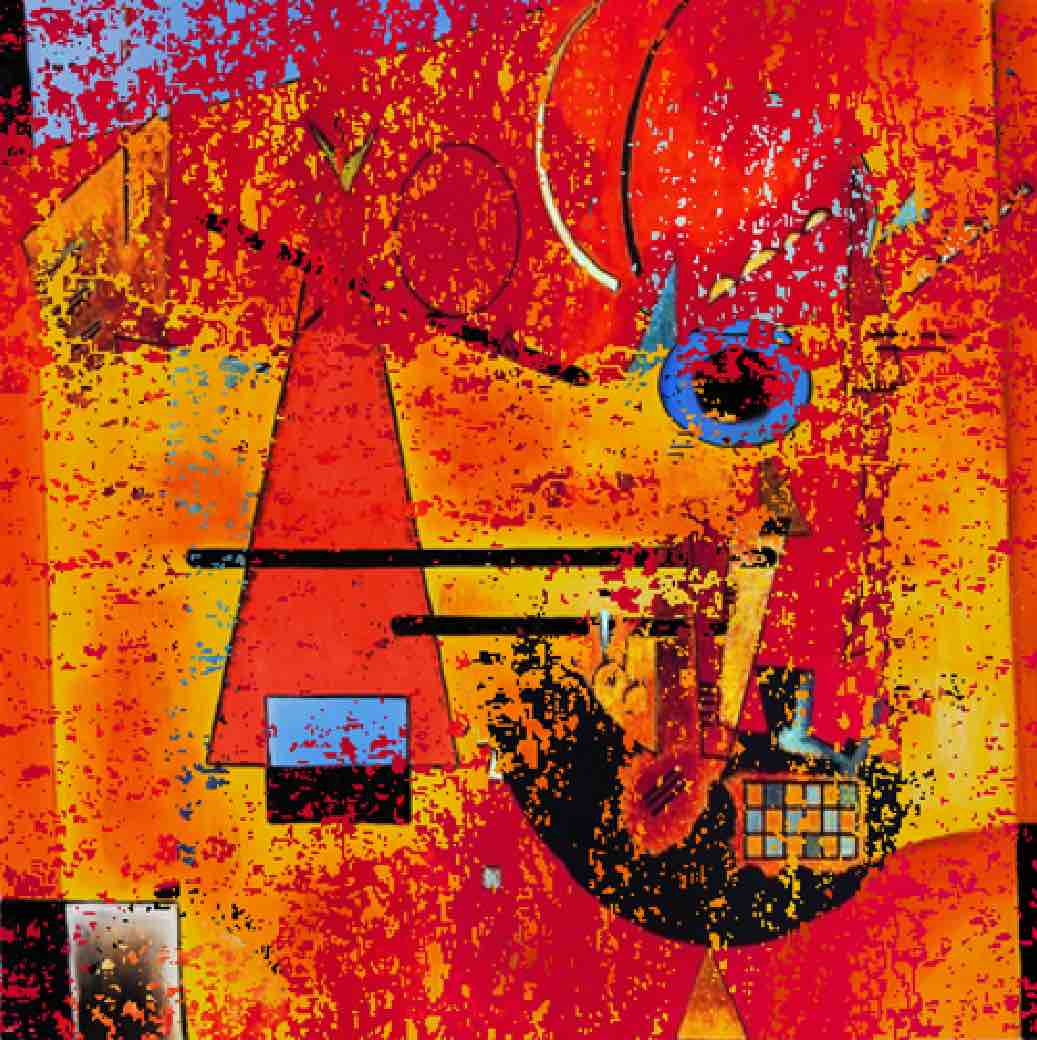}} & 
\subcaptionbox*{}{\includegraphics[width = 1.225in]{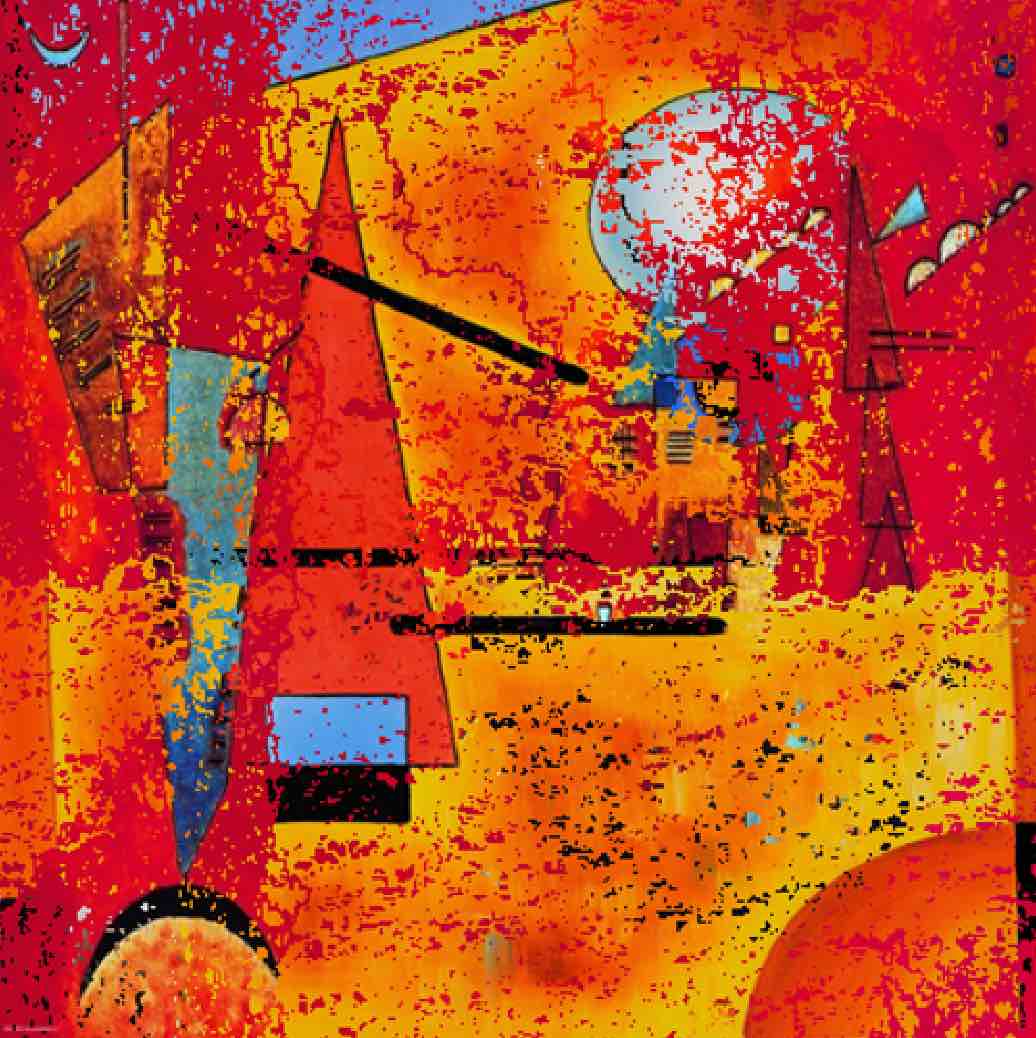}} \\

\end{tabular}
\caption{Quasi-random animation with $2$ Agents. 
Top row shows symmetric sequences (images after~\,3\,600\,000, 3\,900\,000, 4\,000\,000 and 4\,900\,000 steps).
Bottom row shows asymmetric sequences (images after~\,300\,000, 500\,000, 1\,600\,000 and 3\,700\,000 steps)}
\label{TWO_Walkers_Animation}
\end{figure*}

We use $2$ agents starting at positions $[m/4, n/4]$ and $[3m/4, 3n/4]$ and images from Figure~\ref{source_and_targets}. Image (c) is used as the starting image and as image $I^1$ for the first agent, and image (d) is used image as $I^2$ for the second agent. 
Firstly, we consider the behavior when using the symmetric sequences $S^1$ = (right, down, left, up) and $S^2$ = (up, left, down, right). Figure~\ref{TWO_Walkers_Animation} (top row) shows the results.  
The second agent starts to paint the image in circle appearances according to the symmetric agent movement. The animation obtains a more dynamic character when the regions painted by the agents meet and the agent operate at a "larger radius". This leads to the effect that the agents operate from all direction on the image. It looks as if the agents naturally pervade the image and effortlessly collaborate together. 

We now consider $2$ agents using the asymmetric sequences $S^1$ = (right, down, left, up, right) and $S^2$ = (up, left, down, right, up). Figure~\ref{TWO_Walkers_Animation} (bottom row) shows the results. In contrast to the previous animation, the images obtained are a mixture of the two images of the agents in all  parts of the image. This is due to the first agent painting its image in form of horizontal stripes and the second agent painting its image using vertical strips. As the agents move over the image, their current horizontal (for the first agent) and vertical (for the second agent) position determines which image of the agents shines through in which part at any given point in time of the animation.

\subsubsection{4  Agents}

\begin{figure*}[!t]     
\centering
\begin{tabular}{cccccccc}
\subcaptionbox*{}{\includegraphics[width = 1.225in]{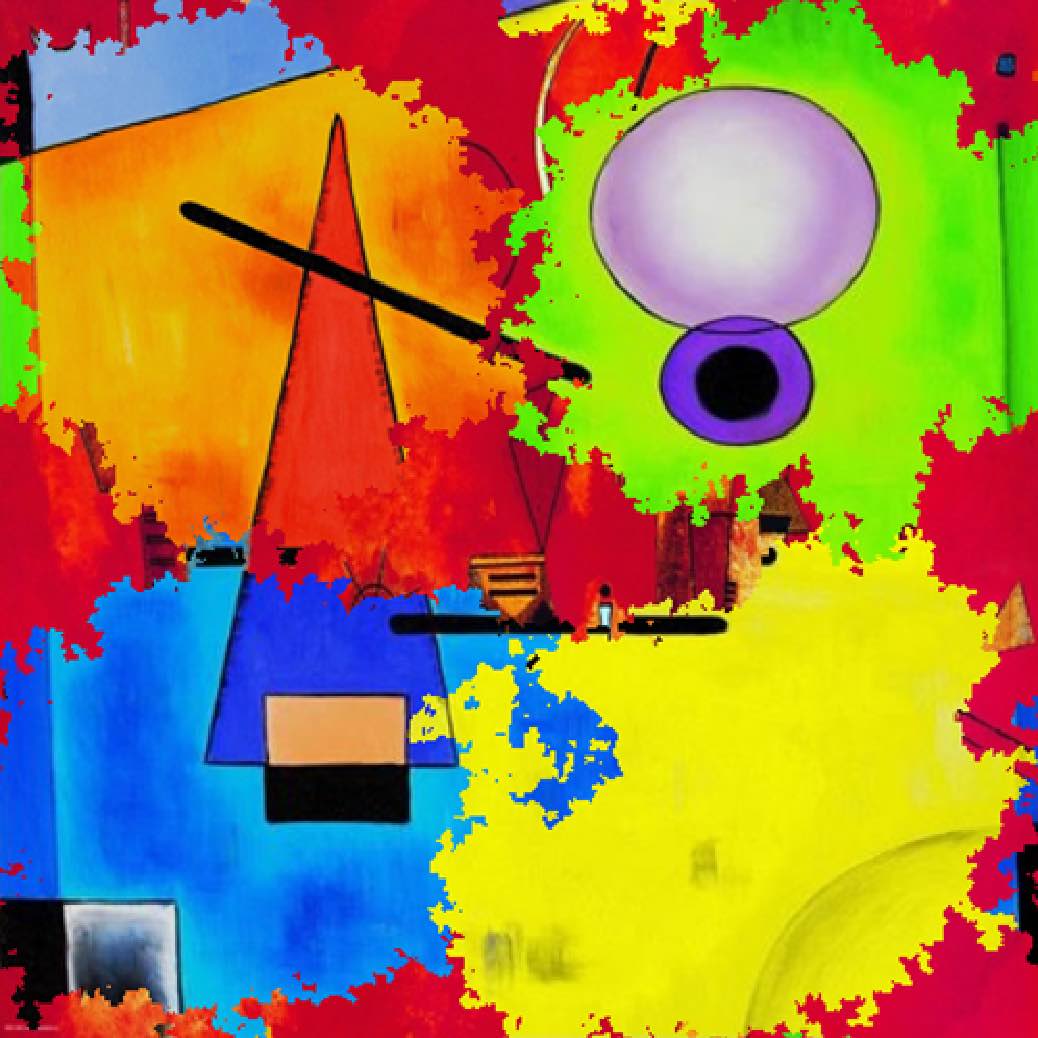}} & 
\subcaptionbox*{}{\includegraphics[width = 1.225in]{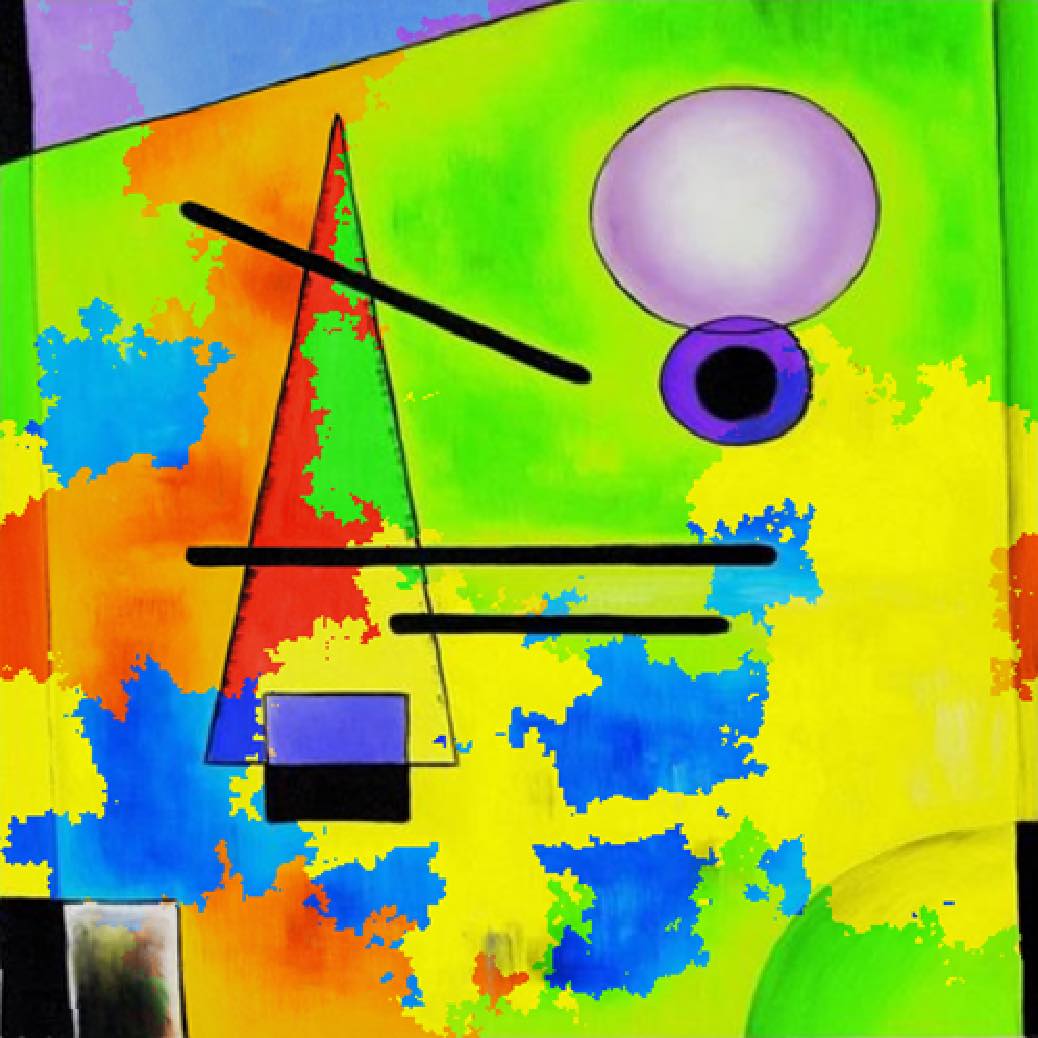}} & 
\subcaptionbox*{}{\includegraphics[width = 1.225in]{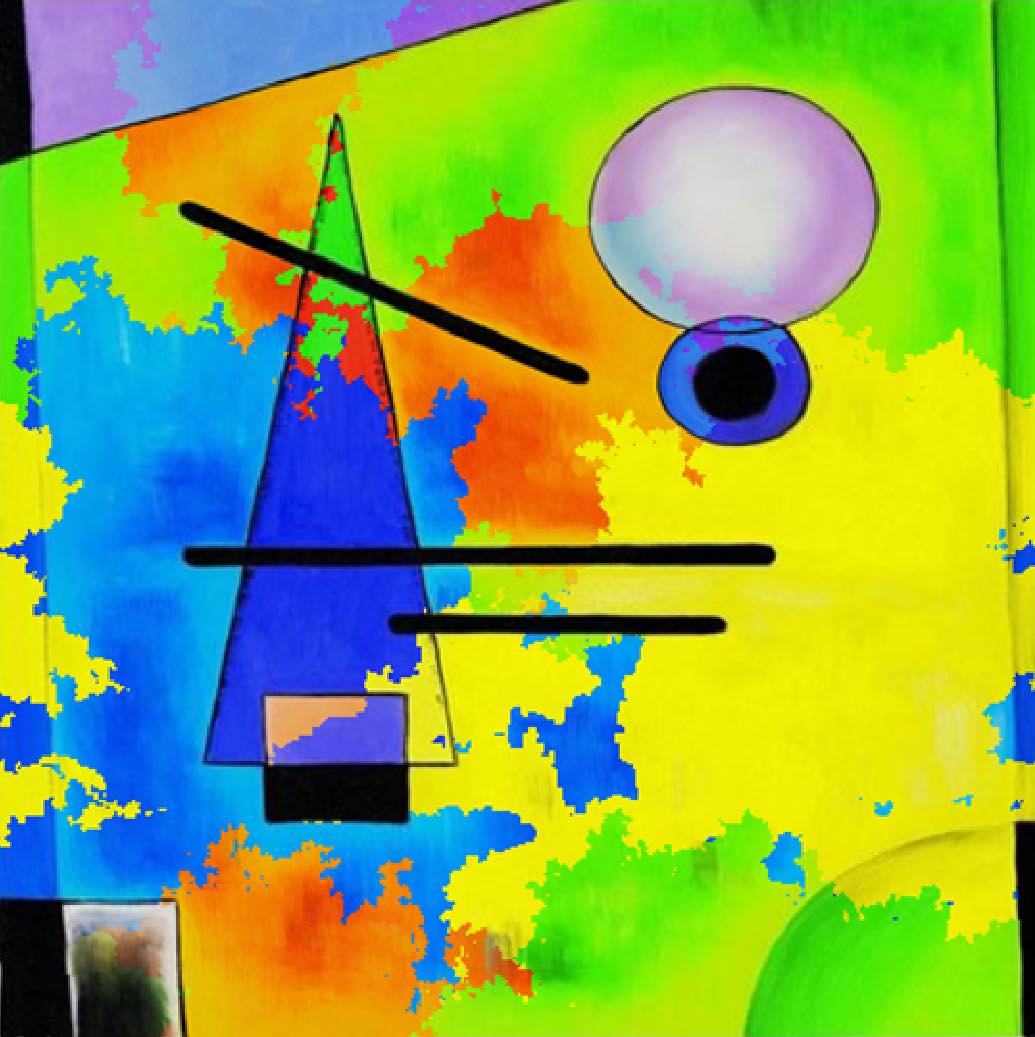}} & 
\subcaptionbox*{}{\includegraphics[width = 1.225in]{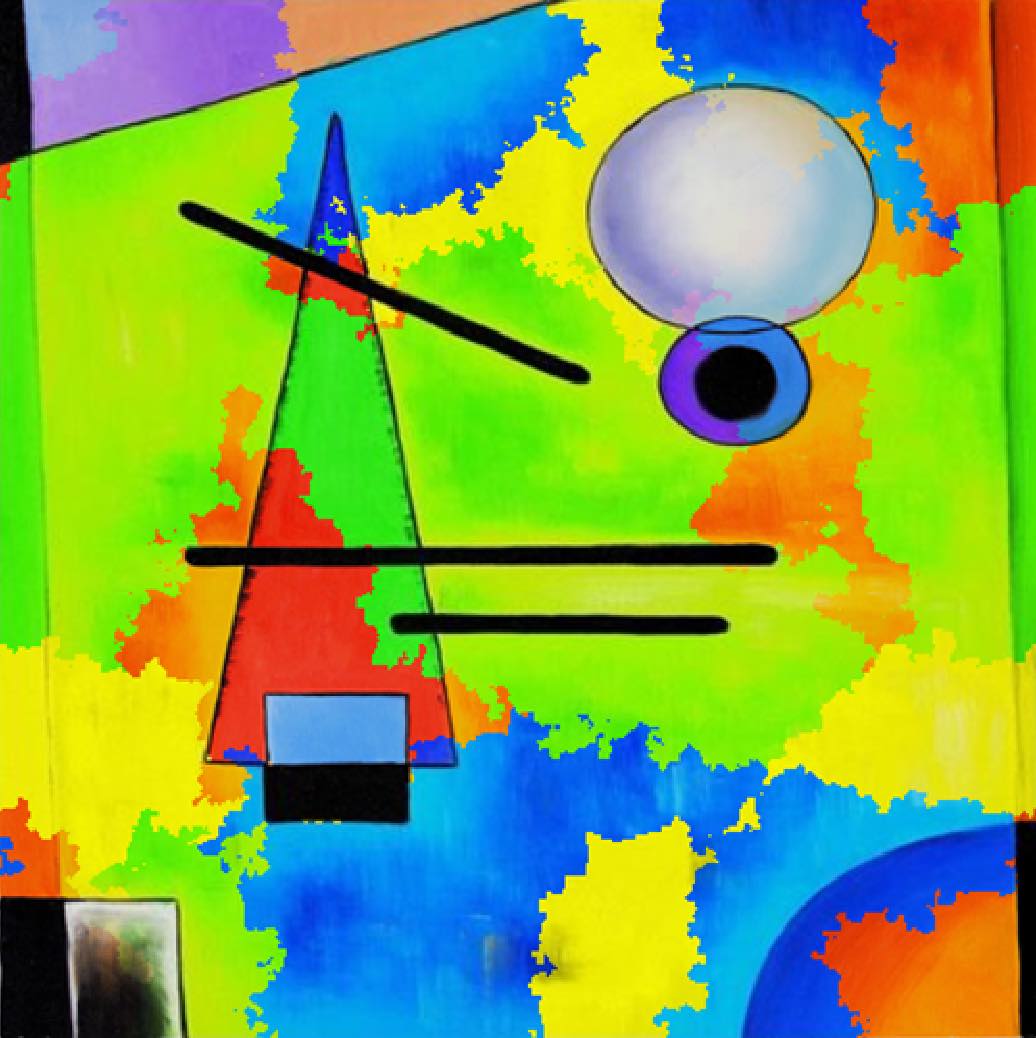}} \\ 
\vspace{-0.3cm}

\subcaptionbox*{}{\includegraphics[width = 1.225in]{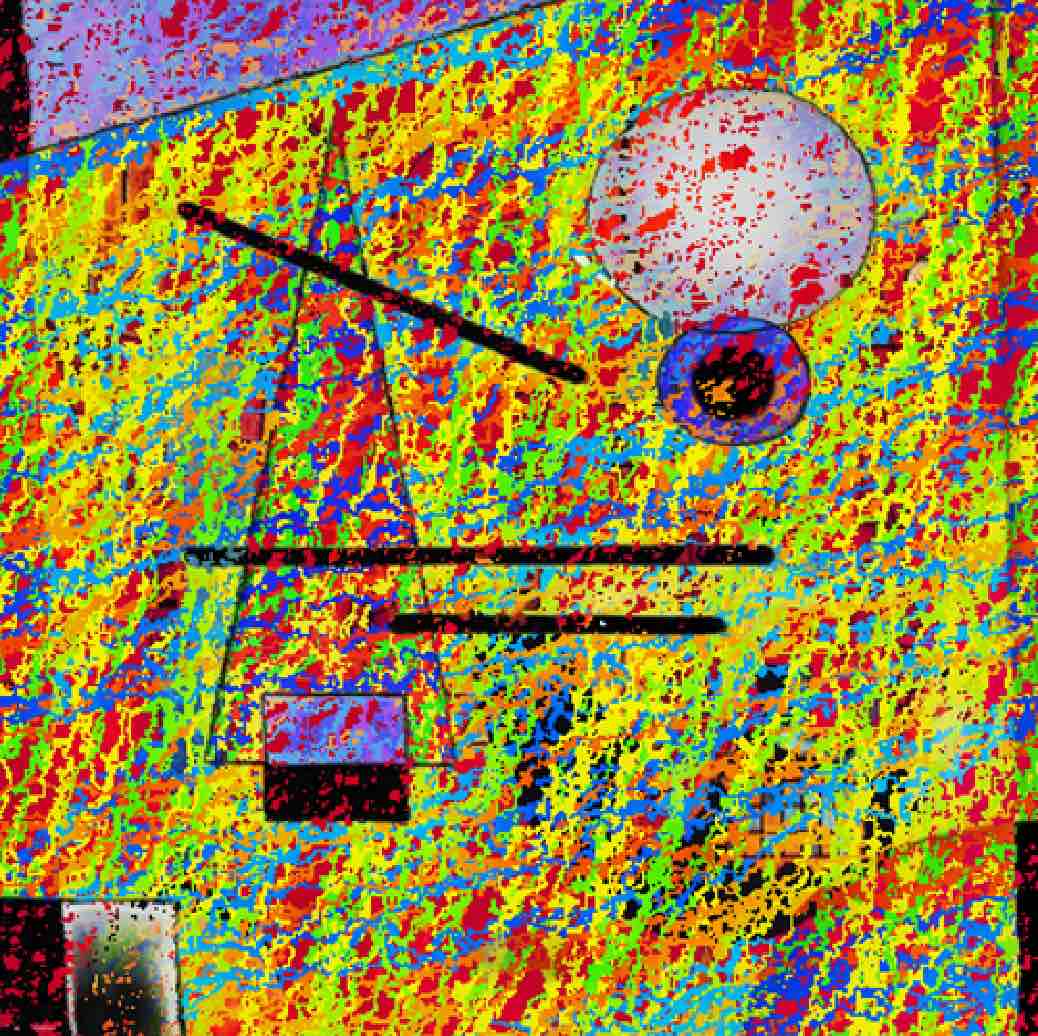}} &  
\subcaptionbox*{}{\includegraphics[width = 1.225in]{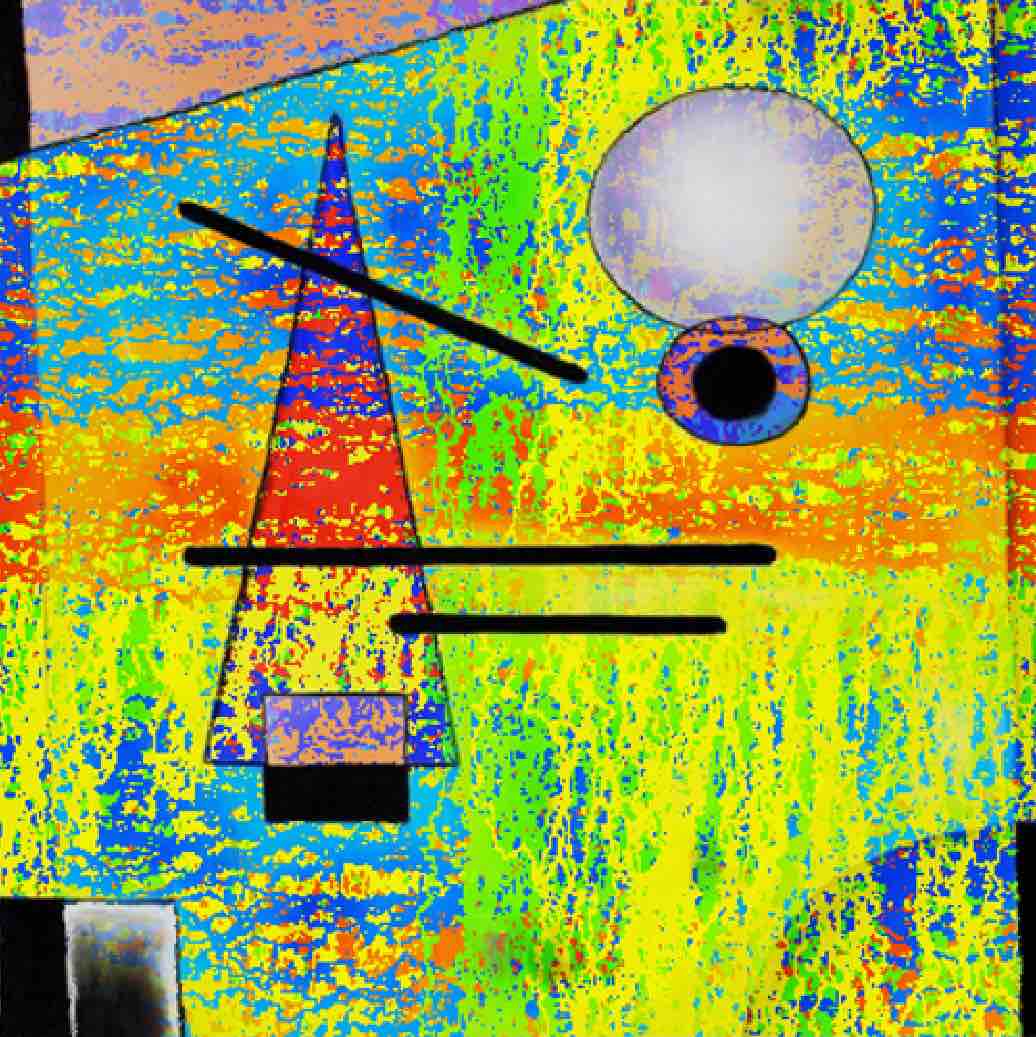}} & 
\subcaptionbox*{}{\includegraphics[width = 1.225in]{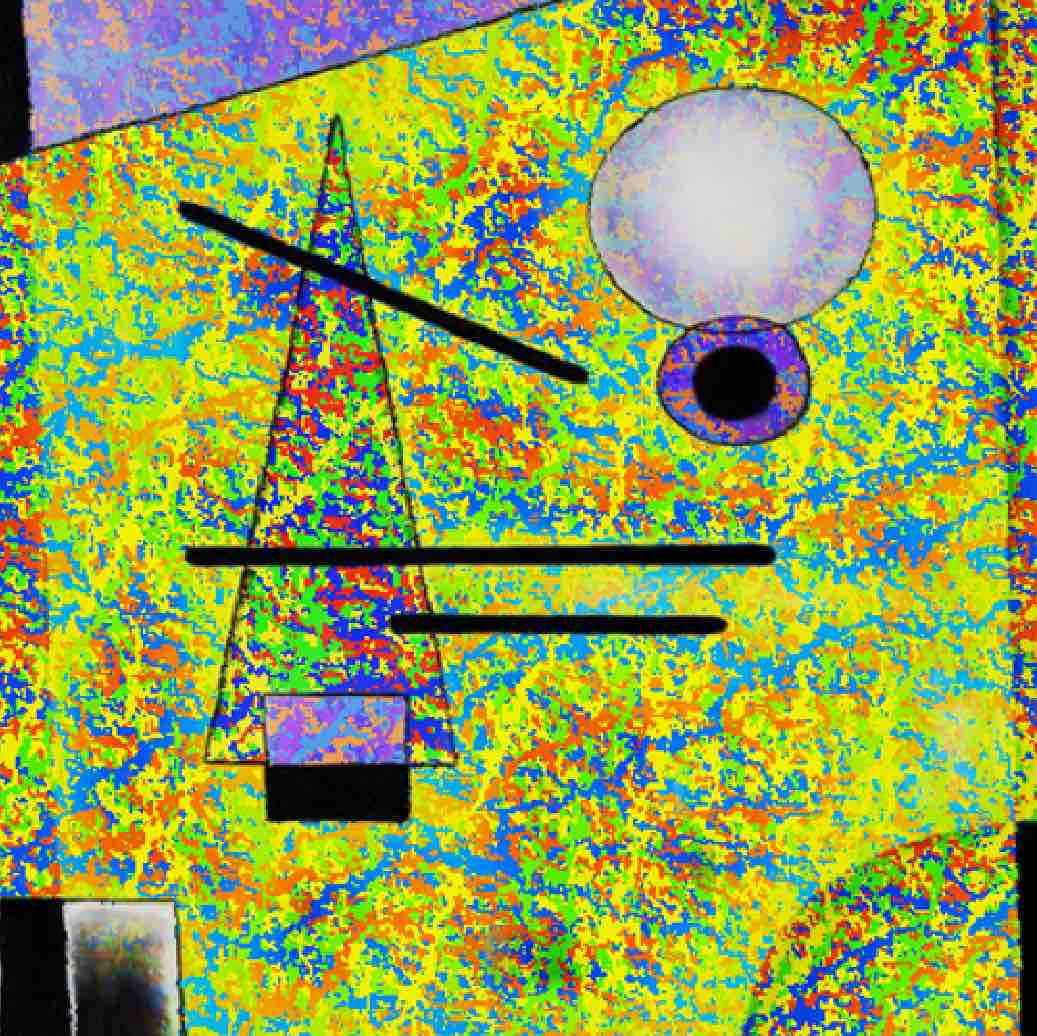}} & 
\subcaptionbox*{}{\includegraphics[width = 1.225in]{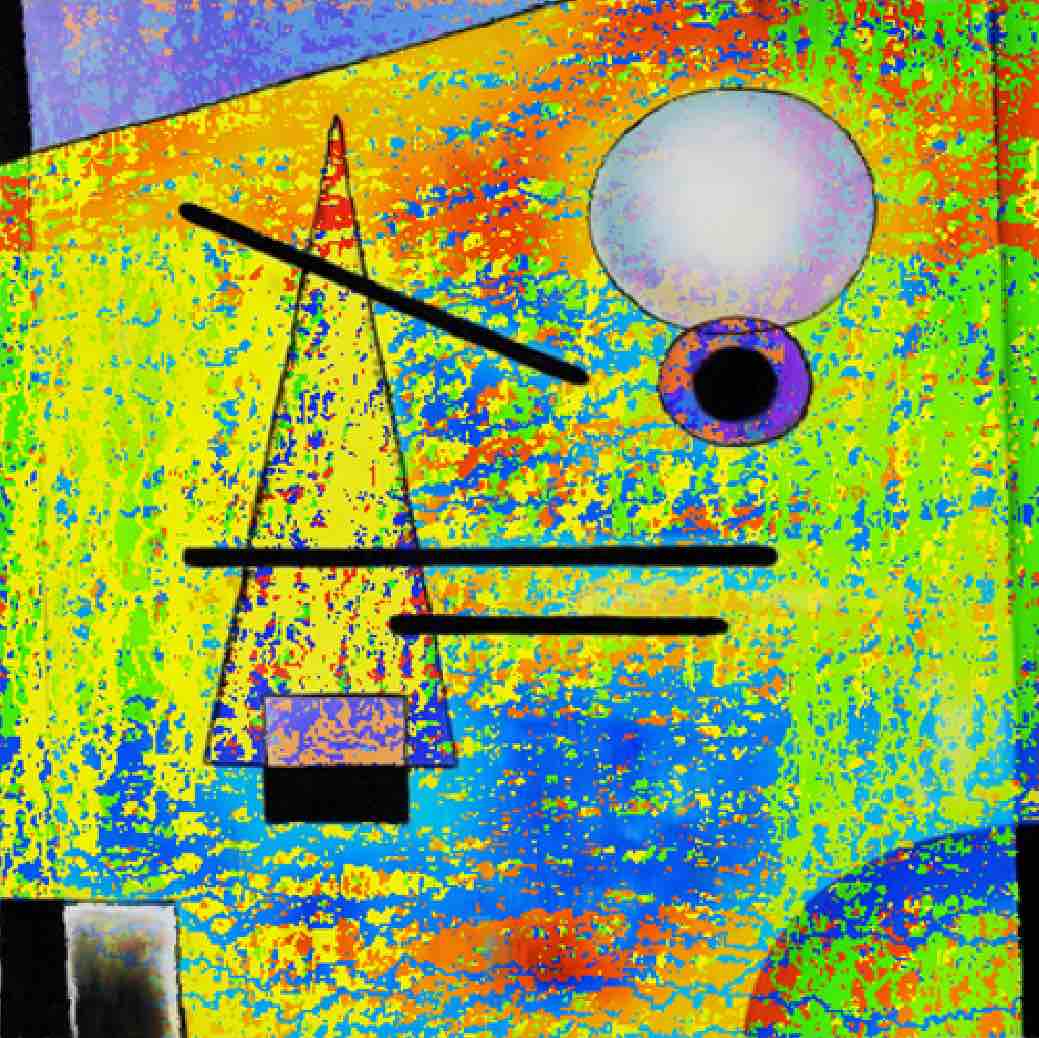}} \\ 
\end{tabular}
\caption{Quasi-random animation with $4$ Agents. 
Top row shows symmetric sequences (images after~\,500\,000, 2\,100\,000, 2\,400\,000 and 4\,200\,000 steps).
Bottom row shows asymmetric sequences (images after\,100\,000, 1\,200\,000, 4\,000\,000 and 4\,300\,000 steps)}
\label{Four_Walkers_longer}
\end{figure*}

We now consider results that are achieved when using $4$ agents.
We obtain $3$ additional images by changing the color spectrum of the image shown in Figure~\ref{source_and_targets}~(c) to blue, green, yellow. 
We use (c) and these $3$ additional images as the images $I^k$ painted by the agents. This means that we can identify the image of an agent by it's color. The setting allows us to design animations where the underlying images have the same structure and only differ in terms of their color spectrum. The image in Figure~\ref{source_and_targets} (d) is chosen as the starting image.  
The $4$ agents start at positions $[m/4, m/4]$, $[m/4, 3n/4]$, $[3m/4, n/4]$ and $[3m/4, 3n/4]$, respectively.

For the symmetric experiment, we use the sequences $S^1=S^3$ = (right, down, left, up), and $S^2=S^4$ = (up, left, down, right). The results in Figure~\ref{Four_Walkers_longer} (top row) show that the images of the agents  appear in circle-patch styles. It can be observed that parts of the different images painted by the agents appear at different time steps at different parts of the image. Thus, the algorithm allows the agents to paint part of their own image.  The quasi-random image animation produces a visual pleasing result and combines the different parts of all images in a new abstract composition.

We now consider $4$ agents using the asymmetric sequences $S^1=S^3$ = (right, down, left, up, right), and $S^2=S^4$ = (up, left, down, right, up). Figure~\ref{Four_Walkers_longer} (bottom row) shows the results. In contrast to the symmetric sequences, the different images obtained provide an interesting mixture of the images of the different agents which is due to $2$ agents moving horizontally and $2$ agents moving vertically. There are no parts that can be clearly attribute to one of the agent's image. Instead of this, the animation produces interesting overlapping effects where the dominance of the colors (corresponding the agent's images) in the different parts of the image change over time.